# A Physics-Constrained Data-Driven Approach Based on Locally Convex Reconstruction for Noisy Database

Qizhi He, Jiun-Shyan Chen*

*Department of Structural Engineering, University of California, San Diego, La Jolla, CA, 92093, USA*

## Abstract

Physics-constrained data-driven computing is an emerging hybrid approach that integrates universal physical laws with data-driven models of experimental data for scientific computing. A new data-driven simulation approach coupled with a locally convex reconstruction, termed the local convexity data-driven (LCDD) computing, is proposed to enhance accuracy and robustness against noise and outliers in data sets in the data-driven computing. In this approach, for a given state obtained by the physical simulation, the corresponding optimum experimental solution is sought by projecting the state onto the associated local convex manifold reconstructed based on the nearest experimental data. This learning process of local data structure is less sensitive to noisy data and consequently yields better accuracy. A penalty relaxation is also introduced to recast the local learning solver in the context of non-negative least squares that can be solved effectively. The reproducing kernel approximation with stabilized nodal integration is employed for the solution of the physical manifold to allow reduced stress-strain data at the discrete points for enhanced effectiveness in the LCDD learning solver. Due to the inherent manifold learning properties, LCDD performs well for high-dimensional data sets that are relatively sparse in real-world engineering applications. Numerical tests demonstrated that LCDD enhances nearly one order of accuracy compared to the standard distance-minimization data-driven scheme when dealing with noisy database, and a linear exactness is achieved when local stress-strain relation is linear.

**Keywords**: data-driven computing; locally convex reconstruction; manifold learning; noisy data; local convexity data-driven (LCDD); reproducing kernel (RK) approximation

---

* Correspondence to: Jiun-Shyan Chen, Structural Engineering Department, University of California, San Diego (UCSD), La Jolla, CA 92093, USA. E-mail: js-chen@ucsd.edu

# 1. Introduction

With the proliferation of high-resolution datasets and the significant advances in numerical algorithms, the emerging idea by utilizing both data-driven models and physical models simultaneously to enhance traditional scientific computing and engineering design procedures [1,2] has attracted increasingly attentions. This general approach is usually termed as data-driven modeling [3] or data-driven engineering science. Data-driven modeling has a close connection with the various areas such as statistics, data mining, and machine learning, which allow the extraction of insightful information or the hidden structures from large volumes of data [4] for enhanced scientific computing. The data-driven approaches, such as machine learning [5,6], have been widely applied to computational biological [7] and medical diagnosis [8], material informatics [9,10], and other predictive physics problems [4,11].

Recently, these approaches have been extended to the field of engineering mechanics, such as learning constitutive models in solid mechanics [12–14], surrogate models in fluid mechanics [15–17] and physical models or governing equations purely extracted from the collected data [18–20]. In conjunction with machine learning techniques such as manifold learning [21] or neural networks [22], the recent studies [23–25] offer a new paradigm for data-driven computing for various applications such as design of materials [26]. There is a vast body of literature devoted to these subjects, including the recent developments based on nonlinear dimensionality reduction [24], nonlinear regression, deep learning [27–29], among others.

Nevertheless, pure data-driven methodology in the area of simulation-based engineering sciences (SBES) [30] is ineffective since in many physical systems well-accepted physical laws exist while useful data in SBES are very expensive to acquire [20,31]. Thus, it is imperative to develop data-driven simulation approaches that can leverage the physical principles with limited data for highly complex systems. A solution to develop effective predictive models for complex real-world problems is to combine physics-based models with data-driven techniques under a hybrid computational framework. There are three types of hybrid physics-data approaches, depending on the roles of physics laws and data play in the hybrid model. The first approach enforces known physical constraints into data-driven models [32,33], which can be considered as a data-fit type surrogate model. In the second approach, on the contrary, the existing physical models are enriched by the information learned from data. This general framework can be used for obtaining data-enhanced physical models [34,35], online updating dynamical system in a manner similar to data assimilation [36], or constructing reduced-order model [25,37–40]. The third approach is to apply data-driven models and physical models separately to approximate different aspects of the physical system and be connected consistently to perform numerical simulation.

The third class of methods is particularly attractive for computational mechanics because it preserves the well-accepted physical laws under the variational framework and the prerequisite of big data is much relaxed. In contrast, most other data-driven methods for solid or fluid simulation directly construct machine learning surrogates to link the input-output relation with approximated physics laws [17, 24, 33] or with the constitutive models replaced by supervised learning models such as neural networks [12-14, 27], which could over-parameterize the material relation and lead to numerical instability. In addition, the setting of training and architecture hyperparameters for neural networks is not straightforward.

Under the framework of the third approach mentioned above, Kirchdoerfer and Ortiz [41–43] have proposed a material model-free data-driven method, so called distance-minimizing data-driven computing (DMDD), for modeling elasticity problems. This data-driven method enforces equilibrium and compatibility and directly utilizes the material database, e.g. stress and strain data, under a modified variational framework, aiming to eschew the empirical models that inevitably involve incomplete experimental information [41,42] and the process of material parameter identification [44–46] that remains numerically intractable. In DMDD, the data-driven problem is solved by minimizing the distance between the computed physical solutions (i.e. the set of equilibrium admissible stress and kinematically admissible strain jointly) and a given set of experimental data under a proper energy norm. A similar idea was proposed by Ibañez et al. [47] where manifold learning techniques are applied to material database to construct the tangent stiffness approximation of constitutive relation, with which the convergent solution could be attained by using directional search solvers [48,49]. In these methods, the data selection or automated machine learning techniques on material data are carried out during the computation of the associated initial-boundary-value problem, thus bypassing the traditional construction of constitutive models. Again, these methods fall into the third class of the data-driven approaches discussed above, and they are usually defined as data-driven computational mechanics (DDCM) [41,47]. This data-driven paradigm has been recently extended to dynamics [43], nonlinear elasticity [35,50,51,82], material identification [52], and data completion [53]. Overall, the key idea of the above mentioned methods is to seek the intersection of the hidden constitutive (material) manifold represented by experimental data and the physical manifold by using iterative processes with appropriate search directions, as shown in **Fig. 1**.

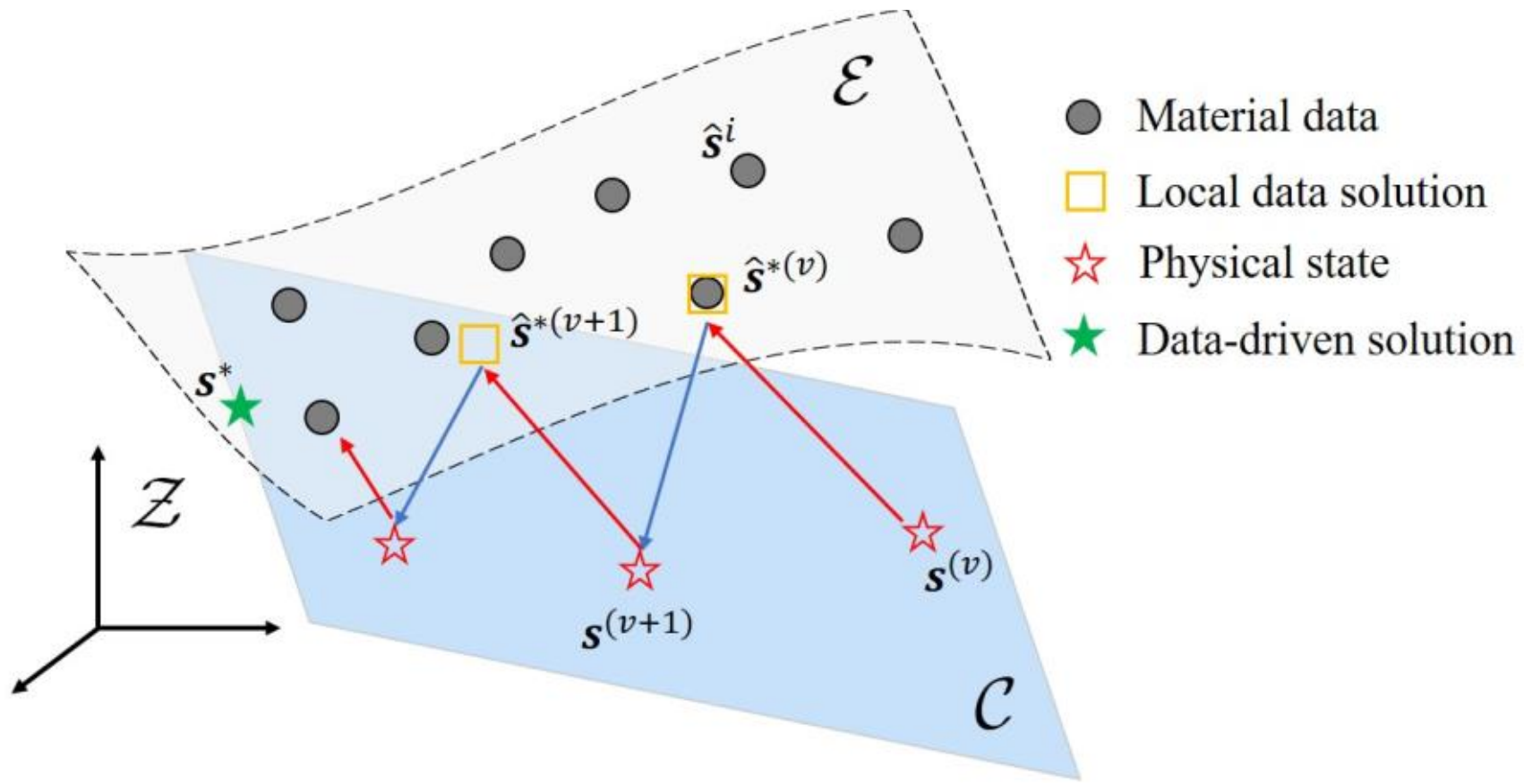

**Fig. 1.** Schematics of data-driven computing for predictive modeling, where a set of data points $\{\hat{s}^i, i = 1, 2, ...\}$ are given to represent the material behavior, $\mathcal{E}$ is an imaginary manifold of the material database representing the underlying constitutive relation, and $\mathcal{C}$ is the physical manifold of admissible stress and strain states $.\, \boldsymbol{s} = (\boldsymbol{\varepsilon}, \boldsymbol{\sigma})$ . satisfying equilibrium and compatibility. The data-driven solution $\boldsymbol{s}^*$ is solved by a fixed point iteration that searches the physical states $\boldsymbol{s}^{(v)} \in \mathcal{C}$ and the local data solutions $\hat{\boldsymbol{s}}^{*(v)} \in \mathcal{E}$ via iterative projections, where the subscript '$v$' is the step indicator. In this study, only stresses and strains are considered in the material database.

Despite the major advancement made in the field, it remains challenging in dealing with noisy and sparse data [1]. The standard DMDD paradigm [41] is shown sensitive to noisy data and outliers [42,54], while the approaches based on manifold learning [47] or local regression [55] may fail to converge due to the over-relaxed manifold constrsuction and lack of convexity. To enhance robustness, the DMDD approach was extended to the max-ent data driven computing [42], which utilizes entropy estimation to analyze the statistics information of data. However, as simulated annealing algorithms are used to solve the resulting data-driven minimization problem, the solution procedures become computationally intensive. Alternatively, Ayensa-Jimenez et al.

[54] proposed to consider data uncertainty by explicitly incorporating statistical quantities into the standard DMDD approach and defined a stochastic analogous problem. In this approach, the expected value and variance-covariance matrix need to be estimated and the local data structure is still not taken into consideration and hence it becomes ineffective in dealing with high-dimensional dataset.

In this paper, we propose a novel data-driven approach which utilizes the intrinsic local data structure to enhance accuracy and robustness against noisy data while computationally feasible. By assuming that each data point and its neighbors lie on or close to a locally linear patch of the manifold, the proposed approach approximates the underlying constitutive manifold near the physically admissible state by locally constructing a convex envelop based on the associated experimental neighbor data points. As a result, the proposed approach can utilize the local data structure without explicitly constructing the local manifold or regression models needed in the approaches [47,54,55] utilizing the associated tangent spaces for data-driven iterations. With this locally convex construction, the solution space for searching optimum local data is regularized onto a bounded, continuous, and convex subset (polytope) for enhanced robustness and convergence stability in data-driven computing. The proposed approach is, thus, referred to as local convexity data-driven (LCDD) computing. In this approach, a cluster of experimental data associated with the physical solution (e.g., the pair of strains and stresses) is first identified by the *k-nearest neighbor* (*k*-NN) algorithm, and the optimum data solution is searched within the associated locally convex hull instead of the discrete material set. To solve this local search problem efficiently, we recast the approach into a *non-negative least squares* (NNLS) problem [56] by introducing the invariance constraint into the objectivity function. Because of the inherited manifold learning capacity in the NNLS solvers, the proposed LCDD permits the locally linear approximation for the underlying material manifold, which means that LCDD could reproduce the solutions given by the classical model-based simulation if the constitutive relation represents a locally linear pattern.

On the other hand, LCDD can be viewed as a generalization of DMDD by equipping a suitable manifold learning technique that naturally takes the local data information into account and retains a simple computing framework. In the solution phase on the physical manifold, a constrained minimization problem is solved by introducing a reproducing kernel approximation (RK) [57,58] in conjunction with a stabilized conforming nodal integration (SCNI) [59] such that the displacements, stresses and strains are computed at the nodal points. This approach significantly reduces the needed search for the optimal stress-strain data from the data set. The employment of the RK approximation also introduces higher order smoothness to the solution space of the physical manifold, making it consistent to the continuous and convex solution space of the regularized LCDD learning solver. It is noted that there is no assumption of isotropy and homogeneity in the proposed LCDD data-driven computational framework. The learning algorithm can identify the intrinsic properties of the given dataset. In this study, we only consider the modeling of homogeneous material, that is, the same dataset is used to characterize the material behavior at every evaluation point over the domain.

The objective of the present work is to study the main issues of data-driven approaches when dealing with noisy data in high-dimensional space. The paper is organized as follows. In Section 2, a generalized data-driven computational formalism is reviewed. In Section 3, locally convex reconstruction is introduced and the local manifold learning for data-driven solver formulated under the NNLS framework is presented. Section 4 provides numerical tests of truss structures to demonstrate the effectiveness of LCDD against noisy data. In Section 5, continuum mechanics with elastic solid is considered to assess the accuracy and convergence properties of LCDD when the noisy data is of high-dimensional phase space. Finally, concluding remarks and discussions are given in Section 6.

## 2. Physics-constrained data-driven modeling

In this section, we revisit the methodology of data-driven computational mechanics (DDCM) in [41,47,60] and formulate the associated data-driven problem under the variational framework for distance-minimizing between a physical set and an experimental data set.

The deformation of an elastic solid occupying the domain $\Omega$ bounded by Neumann boundary $\Gamma_t$ and Dirichlet boundary $\Gamma_u$ can be described by two basic laws:

(i) Equilibrium

$$\nabla \cdot \boldsymbol{\sigma} + \boldsymbol{b} = \boldsymbol{0}, \text{ in } \Omega, \tag{1a}$$

$$\boldsymbol{\sigma} \cdot \boldsymbol{n} = \bar{\boldsymbol{t}}, \text{ on } \Gamma_t, \tag{1b}$$

(ii) Compatibility

$$\boldsymbol{\varepsilon} = \frac{1}{2}\left(\nabla \boldsymbol{u} + \nabla \boldsymbol{u}^T\right), \text{ in } \Omega, \tag{2a}$$

$$\boldsymbol{u} = \bar{\boldsymbol{u}}, \text{ on } \Gamma_u, \tag{2b}$$

where $\boldsymbol{\sigma}$, $\boldsymbol{u}$, $\boldsymbol{\varepsilon}$, and $\boldsymbol{b}$ are the Cauchy stress, displacement, strain, and body force, respectively, and $\bar{\boldsymbol{u}}$ and $\bar{\boldsymbol{t}}$ are the prescribed displacement and the applied traction on $\Gamma_u$ and $\Gamma_t$, respectively.

Traditionally, to solve a boundary value problem (BVP) in (1) and (2), the constitutive law, e.g. $\boldsymbol{\sigma} = \boldsymbol{\sigma}(\boldsymbol{\varepsilon})$, is required to relate $\boldsymbol{\sigma}$ and $\boldsymbol{\varepsilon}$. In data-driven computing, the equilibrium and compatibility equations in (1) and (2) are solved numerically, while the constitutive relation is given by a set of experimental data, $\mathbb{E} = \{(\hat{\boldsymbol{\varepsilon}}^i, \hat{\boldsymbol{\sigma}}^i), i = 1, ..., p\}$, where $p$ is the number of measurement data collected from experiments.

**Remark 2.1** *For modeling homogeneous material, the same dataset* $\mathbb{E}$ *is used to characterize the material behavior at every evaluation point over the domain. Note that this data-driven computing can be applied to heterogeneous materials if space-dependent databases are available. In this approach, there is no predefined material models or material parameters that need to be identified, which makes data-driven computing different from the classical simulation methods and material identification problems.*

It is convenient to introduce the notion of phase space $\mathcal{Z}$ as the space of the strain-stress pairs $(\boldsymbol{\varepsilon},\boldsymbol{\sigma})$, and denote $\mathcal{C}$ as the admissible set for elements $(\boldsymbol{\varepsilon},\boldsymbol{\sigma})\in\mathcal{Z}$ that satisfy the physical constraints in (1) and (2), which is also called the physical manifold. Ideally, the data-driven solution is the intersection of the global data set $\mathbb{E}_{em}$ and the physical manifold set $\mathcal{C}$, i.e. $\mathbb{E}_{em}\bigcap\mathcal{C}$, where $\mathbb{E}_{em}=\mathbb{E}\times\ldots\times\mathbb{E}\subset\mathcal{Z}$ denotes the ensemble of the experimental set $\mathbb{E}$ over $\Omega$. Since $\mathbb{E}$ consists of a finite set of discrete data points which could lead to non-existence of the intersection $\mathbb{E}_{em}\bigcap\mathcal{C}$, a distance-minimizing relaxation is usually employed.

### *2.1. Distance-minimizing data-driven problem*

Data-driven computing [41,47] and data-enabled applications, such as dynamics data-driven application systems (DDDAS) [1] and parameter identification for pre-defined material models [44–46], introduce distance-minimization between the simulation data and the measurement data. The main difference between these approaches is that data-driven computing is a forward problem while parameter identification is an inverse problem for material calibration. We refer interested readers to the literatures for more details of parameter identification [44–46]. Data-driven computing can be stated as one of the following double-minimization problems:

$$\min_{(\hat{\boldsymbol{\varepsilon}},\hat{\boldsymbol{\sigma}})\in\mathbb{E}_{em}}\ \min_{(\boldsymbol{u},\boldsymbol{\sigma})\in\mathcal{C}_u\times\mathcal{C}_\sigma}\mathcal{H}(\boldsymbol{u},\boldsymbol{\sigma},\hat{\boldsymbol{\varepsilon}},\hat{\boldsymbol{\sigma}})\ \text{ or }\ \min_{(\boldsymbol{u},\boldsymbol{\sigma})\in\mathcal{C}_u\times\mathcal{C}_\sigma}\ \min_{(\hat{\boldsymbol{\varepsilon}},\hat{\boldsymbol{\sigma}})\in\mathbb{E}_{em}}\mathcal{H}(\boldsymbol{u},\boldsymbol{\sigma},\hat{\boldsymbol{\varepsilon}},\hat{\boldsymbol{\sigma}}) \tag{3}$$

where $\mathcal{H}$ is a given functional to define a distance measure, which is to be elaborated in the next section, and $\mathcal{C}_\sigma$ and $\mathcal{C}_u$ denote the sets of *equilibrium admissible* stress fields and *kinematically admissible* displacement fields, respectively, i.e.,

$$\mathcal{C}_\sigma=\left\{\boldsymbol{\tau}\in\mathcal{V}_\sigma \mid \nabla\cdot\boldsymbol{\tau}+\boldsymbol{b}=\boldsymbol{0}\text{ in }\Omega,\text{ and }\boldsymbol{\tau}\cdot\boldsymbol{n}=\bar{\boldsymbol{t}}\text{ on }\Gamma_t\right\}, \tag{4a}$$

$$\mathcal{C}_u=\left\{\boldsymbol{v}\in\mathcal{V}_u \mid \boldsymbol{v}=\bar{\boldsymbol{u}}\text{ on }\Gamma_u\right\}, \tag{4b}$$

in which $\mathcal{V}_\sigma=[L^2(\Omega)]^6$ is the symmetric stress space, and $\mathcal{V}_u=[H^1(\Omega)]^3$ is the displacement space. Then the physical manifold set is defined as

$$\mathcal{C}=\left\{(\boldsymbol{\varepsilon}[\boldsymbol{u}],\boldsymbol{\sigma})\mid \boldsymbol{u}\in\mathcal{V}_u,\boldsymbol{\sigma}\in\mathcal{V}_\sigma\right\}. \tag{4c}$$

Note that strain $\boldsymbol{\varepsilon}$ is obtained from the displacement $\boldsymbol{u} \in \mathcal{C}_u$ using (2a), which is denoted by $\boldsymbol{\varepsilon} = \boldsymbol{\varepsilon}[\boldsymbol{u}]$.

Corresponding to the strain-stress state $(\boldsymbol{\varepsilon}, \boldsymbol{\sigma}) \in \mathcal{C}$ obtained from the physical manifold $\mathcal{C}$, $(\hat{\boldsymbol{\varepsilon}}, \hat{\boldsymbol{\sigma}}) \in \mathbb{E}_{em}$ is used to denote the data from the experimental data set $\mathbb{E}_{em}$. As illustrated in **Fig. 1**, the data-driven computing in (3) is to find the state $(\boldsymbol{\varepsilon}, \boldsymbol{\sigma})$ constrained to the physical set $\mathcal{C}$ while closest to the dataset $\mathbb{E}_{em}$ under a certain "distance" measure defined by the functional $\mathcal{H}$, such that the system response is determined directly from the experimental data without specifying any constitutive models.

The data-driven problem in (3) can be decomposed into a two-step problem:

$$\textbf{Global step}: \quad J(\hat{\boldsymbol{\varepsilon}}, \hat{\boldsymbol{\sigma}}) = \min_{(\boldsymbol{u}, \boldsymbol{\sigma}) \in \mathcal{C}_u \times \mathcal{C}_\sigma} \mathcal{H}(\boldsymbol{u}, \boldsymbol{\sigma}, \hat{\boldsymbol{\varepsilon}}, \hat{\boldsymbol{\sigma}}), \tag{5a}$$

$$\textbf{Local step}: \quad (\hat{\boldsymbol{\varepsilon}}^*, \hat{\boldsymbol{\sigma}}^*) = \arg \min_{(\hat{\boldsymbol{\varepsilon}}, \hat{\boldsymbol{\sigma}}) \in \mathbb{E}_{em}} J(\hat{\boldsymbol{\varepsilon}}, \hat{\boldsymbol{\sigma}}) \tag{5b}$$

where $(\hat{\boldsymbol{\varepsilon}}^*, \hat{\boldsymbol{\sigma}}^*)$ is the optimum experimental point closest to the computed state $(\boldsymbol{\varepsilon}, \boldsymbol{\sigma})$ given in (5 a). From an optimization perspective, the solution procedures of this data-driven problem involve an alternate-direction search where a minimization with respect to $(\boldsymbol{u}, \boldsymbol{\sigma})$ is followed by a minimization with respect to $(\hat{\boldsymbol{\varepsilon}}, \hat{\boldsymbol{\sigma}})$, denoted as a *global step* and a *local step*, respectively.

Compared to the problem setting in material parameter identification [45], the data-driven computing in (5) does not rely on any pre-assumed elasticity tensor to relate $\boldsymbol{\varepsilon}$ and $\boldsymbol{\sigma}$. Instead, it iteratively searches a representative stress-strain pair from the experimental dataset for performing simulation.

### *2.2. Data-driven solver*

The norm $||\bullet||_{\mathcal{Z}}$ associated to the phase space $\mathcal{Z}$ has been defined as a combination of the energy-like and complementary energy-like functional [41] as follows:

$$||(\boldsymbol{\varepsilon}, \boldsymbol{\sigma})||_{\mathcal{Z}}^2 = \frac{1}{2} \int_\Omega \boldsymbol{\varepsilon} : \boldsymbol{M}^\varepsilon : \boldsymbol{\varepsilon} + \boldsymbol{\sigma} : \boldsymbol{M}^\sigma : \boldsymbol{\sigma} \mathrm{d}\Omega, \tag{6}$$

where $\boldsymbol{M}^{\varepsilon}$ and $\boldsymbol{M}^{\sigma}$ are two tensors to balance the contribution of the strain and stress data measured in different physical units.

For numerical implementation, the state variables $(\boldsymbol{\varepsilon}, \boldsymbol{\sigma})$ are computed at integration points $(\boldsymbol{\varepsilon}_\alpha, \boldsymbol{\sigma}_\alpha) \equiv (\boldsymbol{\varepsilon}(\boldsymbol{x}_\alpha), \boldsymbol{\sigma}(\boldsymbol{x}_\alpha)) \in \mathbb{R}^q \times \mathbb{R}^q$, where $\{\boldsymbol{x}_\alpha\}_{\alpha=1}^m$ are the coordinates of the $m$ integration points (i.e., stress-strain evaluation points) and $q$ is the dimension of stress and strain. As such, we denote $\{(\boldsymbol{\varepsilon}_\alpha, \boldsymbol{\sigma}_\alpha)\}_{\alpha=1}^m \in \mathcal{Z}^h$ , where $\mathcal{Z}^h \subset \mathcal{Z}$ is the discrete counterpart of the phase space. Correspondingly, the distance minimization in the local step searches for the local data solution $(\hat{\boldsymbol{\varepsilon}}_\alpha, \hat{\boldsymbol{\sigma}}_\alpha) \in \mathbb{E}$ at every integration point $\boldsymbol{x}_\alpha$ , $\alpha = 1,...,m$. In the subsequent discussion, we define $\boldsymbol{s}_\alpha = [\boldsymbol{\varepsilon}_\alpha^T \ \boldsymbol{\sigma}_\alpha^T]^T \in \mathbb{R}^{2q}$ and $\hat{\boldsymbol{s}}_\alpha = [\hat{\boldsymbol{\varepsilon}}_\alpha^T \ \hat{\boldsymbol{\sigma}}_\alpha^T]^T \in \mathbb{R}^{2q}$ as the computational and experimental strain-stress pairs, respectively, in the local phase space.

A functional $\mathcal{H}$ defined as the discrete form of (6) to measure the distance between $\{(\boldsymbol{\varepsilon}_\alpha, \boldsymbol{\sigma}_\alpha)\}_{\alpha=1}^m$ and $\{(\hat{\boldsymbol{\varepsilon}}_\alpha, \hat{\boldsymbol{\sigma}}_\alpha)\}_{\alpha=1}^m$ is given as

$$\mathcal{H}(\boldsymbol{u}, \boldsymbol{\sigma}, \hat{\boldsymbol{\varepsilon}}, \hat{\boldsymbol{\sigma}}) = \| (\boldsymbol{\varepsilon} - \hat{\boldsymbol{\varepsilon}}, \boldsymbol{\sigma} - \hat{\boldsymbol{\sigma}}) \|_{\mathcal{Z}}^2 \approx \sum_{\alpha=1}^{m} d^2(\boldsymbol{s}_\alpha, \hat{\boldsymbol{s}}_\alpha) V_\alpha, \tag{7}$$

where $\{V_\alpha\}_{\alpha=1}^m$ are the quadrature weights associated with the $m$ integration points, and

$$d(\boldsymbol{s}_\alpha, \hat{\boldsymbol{s}}_\alpha) = (d^{\varepsilon 2}(\boldsymbol{\varepsilon}_\alpha, \hat{\boldsymbol{\varepsilon}}_\alpha) + d^{\sigma 2}(\boldsymbol{\sigma}_\alpha, \hat{\boldsymbol{\sigma}}_\alpha))^{1/2}, \tag{8}$$

where

$$\begin{aligned} d^{\varepsilon}(\boldsymbol{\varepsilon}_\alpha, \hat{\boldsymbol{\varepsilon}}_\alpha) &= \left(1/2\, (\boldsymbol{\varepsilon}_\alpha - \hat{\boldsymbol{\varepsilon}}_\alpha)^T \boldsymbol{M}_\alpha^{\varepsilon} (\boldsymbol{\varepsilon}_\alpha - \hat{\boldsymbol{\varepsilon}}_\alpha)\right)^{1/2}, \\ d^{\sigma}(\boldsymbol{\sigma}_\alpha, \hat{\boldsymbol{\sigma}}_\alpha) &= \left(1/2\, (\boldsymbol{\sigma}_\alpha - \hat{\boldsymbol{\sigma}}_\alpha)^T \boldsymbol{M}_\alpha^{\sigma} (\boldsymbol{\sigma}_\alpha - \hat{\boldsymbol{\sigma}}_\alpha)\right)^{1/2}. \end{aligned} \tag{9}$$

Here $\boldsymbol{M}_\alpha^{\varepsilon} \in \mathbb{R}^{q \times q}$ and $\boldsymbol{M}_\alpha^{\sigma} \in \mathbb{R}^{q \times q}$ are symmetric and positive-definite coefficient matrices for multivariate distance measures, and usually $\boldsymbol{M}_\alpha^{\sigma} = \left(\boldsymbol{M}_\alpha^{\varepsilon}\right)^{-1}$ . One approach for selecting the coefficient matrices is by computing the covariance of the material data set and using the so-called Mahalanobis distance for multivariate data, as proposed in [54]. Investigating the effect of coefficient matrices is out of the scope of this study. Numerical examples show that by using the proposed locally convex construction scheme with a coefficient matrix representing linear elasticity, which can be extracted from stress-strain dataset at small strain range, satisfactory data-driven results are achieved.

### *2.3.1. Global step of data-driven solver*

The global step of the data-driven problem (5a) is reformulated as:

$$\begin{aligned} &\min_{\boldsymbol{u}\in\mathcal{C}_u,\boldsymbol{\sigma}\in\mathcal{V}_\sigma} \mathcal{H}(\boldsymbol{u},\boldsymbol{\sigma},\hat{\boldsymbol{\varepsilon}},\hat{\boldsymbol{\sigma}}), \\ &\text{subject to:} \quad \text{div}\boldsymbol{\sigma}+\boldsymbol{b}=\boldsymbol{0} \quad \text{in } \Omega, \\ &\qquad\qquad\quad \boldsymbol{\sigma}\cdot\boldsymbol{n}=\bar{\boldsymbol{t}} \qquad \text{on } \Gamma_t. \end{aligned} \tag{10}$$

This global step searches for the physically admissible state $\boldsymbol{s}=(\boldsymbol{\varepsilon}[\boldsymbol{u}],\boldsymbol{\sigma})\in\mathcal{C}$ closest to a given experimental data $\hat{\boldsymbol{s}}=(\hat{\boldsymbol{\varepsilon}},\hat{\boldsymbol{\sigma}})$ by means of Lagrange multipliers

$$\mathcal{L}_{DD}(\boldsymbol{u},\boldsymbol{\sigma},\boldsymbol{\lambda},\boldsymbol{\eta})=\mathcal{H}(\boldsymbol{u},\boldsymbol{\sigma},\hat{\boldsymbol{\varepsilon}},\hat{\boldsymbol{\sigma}})+\int_\Omega \boldsymbol{\lambda}\cdot(\text{div}\boldsymbol{\sigma}+\boldsymbol{b})\text{d}\Omega+\int_{\Gamma_t}\boldsymbol{\eta}\cdot(\boldsymbol{\sigma}\cdot\boldsymbol{n}-\bar{\boldsymbol{t}})\text{d}\Gamma, \tag{11}$$

where $\boldsymbol{\lambda}$ and $\boldsymbol{\eta}$ are the Lagrange multipliers in proper function spaces. The Euler-Lagrange equations of (11) reveals $\boldsymbol{\eta}=-\boldsymbol{\lambda}$ on $\Gamma_t$ [61]. Considering the equations (6) - (9), and $\boldsymbol{\varepsilon}=\boldsymbol{\varepsilon}[\boldsymbol{u}]$, the variational form is

$$\begin{aligned} \delta\mathcal{L}_{DD}(\boldsymbol{u},\boldsymbol{\sigma},\boldsymbol{\lambda}) &= \int_\Omega\Big(\delta\boldsymbol{\varepsilon}[\boldsymbol{u}]:\boldsymbol{M}^\varepsilon:(\boldsymbol{\varepsilon}[\boldsymbol{u}]-\hat{\boldsymbol{\varepsilon}})+\delta\boldsymbol{\sigma}:(\boldsymbol{M}^\sigma:(\boldsymbol{\sigma}-\hat{\boldsymbol{\sigma}})\Big)\text{d}\Omega \\ &\quad -\int_\Omega\delta\boldsymbol{\sigma}:\boldsymbol{\varepsilon}[\boldsymbol{\lambda}]\text{d}\Omega+\int_{\Gamma_u}(\delta\boldsymbol{\sigma}\cdot\boldsymbol{n})\cdot\boldsymbol{\lambda}\text{d}\Gamma-\int_\Omega\delta\boldsymbol{\varepsilon}[\boldsymbol{\lambda}]:\boldsymbol{\sigma}\text{d}\Omega \\ &\quad +\int_\Omega\delta\boldsymbol{\lambda}\cdot\boldsymbol{b}\text{d}\Omega+\int_{\Gamma_t}\delta\boldsymbol{\lambda}\cdot\bar{\boldsymbol{t}}\text{d}\Gamma, \end{aligned} \tag{12}$$

where $\boldsymbol{\varepsilon}[\boldsymbol{\lambda}]=1/2(\nabla\boldsymbol{\lambda}+\nabla\boldsymbol{\lambda}^T)$. Consequently, we have

$$\int_\Omega\delta\boldsymbol{\varepsilon}[\boldsymbol{u}]:\boldsymbol{M}^\varepsilon:\boldsymbol{\varepsilon}[\boldsymbol{u}]\text{d}\Omega=\int_\Omega\delta\boldsymbol{\varepsilon}[\boldsymbol{u}]:\boldsymbol{M}^\varepsilon:\hat{\boldsymbol{\varepsilon}}\text{d}\Omega, \tag{13a}$$

$$\int_\Omega\delta\boldsymbol{\varepsilon}[\boldsymbol{\lambda}]:\boldsymbol{\sigma}\text{d}\Omega=\int_\Omega\delta\boldsymbol{\lambda}\cdot\boldsymbol{b}\text{d}\Omega+\int_{\Gamma_t}\delta\boldsymbol{\lambda}\cdot\bar{\boldsymbol{t}}\text{d}\Gamma, \tag{13b}$$

$$\int_\Omega\delta\boldsymbol{\sigma}:(\boldsymbol{M}^\sigma:\boldsymbol{\sigma}-\boldsymbol{\varepsilon}[\boldsymbol{\lambda}])\text{d}\Omega=\int_\Omega\delta\boldsymbol{\sigma}:\boldsymbol{M}^\sigma:\hat{\boldsymbol{\sigma}}\text{d}\Omega. \tag{13c}$$

Note that $\boldsymbol{\lambda}=\boldsymbol{0}$ on $\Gamma_u$ has been introduced. In this study displacement $\boldsymbol{u}$, the Lagrange multipliers $\boldsymbol{\lambda}$ and stress $\boldsymbol{\sigma}$ are approximated by

$$\boldsymbol{u}(\boldsymbol{x}) \approx \boldsymbol{u}^h(\boldsymbol{x}) = \sum_{I=1}^{N} \Psi_I(\boldsymbol{x}) \boldsymbol{d}_I, \tag{14a}$$

$$\boldsymbol{\lambda}(\boldsymbol{x}) \approx \boldsymbol{\lambda}^h(\boldsymbol{x}) = \sum_{I=1}^{N} \Psi_I(\boldsymbol{x}) \boldsymbol{\Lambda}_I, \tag{14b}$$

$$\boldsymbol{\sigma}(\boldsymbol{x}) \approx \boldsymbol{\sigma}^h(\boldsymbol{x}) = \sum_{\alpha=1}^{m} \chi_\alpha(\boldsymbol{x}) \boldsymbol{\sigma}_\alpha, \tag{14c}$$

where $N$ is the number of discretization nodes, $m$ is the number of stress-strain evaluation points at $\boldsymbol{x}_\alpha$, $\{\boldsymbol{d}_I\}_{I=1}^N$ are the nodal displacement vectors, $\{\boldsymbol{\Lambda}_I\}_{I=1}^N$ are the nodal Lagrange multiplier vectors, $\chi_\alpha(\boldsymbol{x})$ is an indicator function such that $\chi_\alpha(\boldsymbol{x}) = 1$ if $\boldsymbol{x} \in \Omega_\alpha$ and $\chi_\alpha(\boldsymbol{x}) = 0$ if $\boldsymbol{x} \notin \Omega_\alpha$, where $\Omega_\alpha$ is the subdomain associated to the integration point $\boldsymbol{x}_\alpha$. Here, we employ $\{\Psi_I\}_{I=1}^N$ the reproducing kernel (RK) shape functions [57,58] constructed using the cubic-B splines kernel function and linear basis functions. The introduction of RK approximation is summarized in Appendix B. Stress in (13c) is discretized by a collocation approach in (14c). Thus, the discrete form of Equation (13c) yields

$$\sum_{\alpha=1}^{m} V_\alpha \delta \boldsymbol{\sigma}_\alpha^T \left( \boldsymbol{M}_\alpha^\sigma \boldsymbol{\sigma}_\alpha - \sum_{I=1}^{N} \boldsymbol{B}_{\alpha I} \boldsymbol{\Lambda}_I \right) = \sum_{\alpha=1}^{m} V_\alpha \delta \boldsymbol{\sigma}_\alpha^T \boldsymbol{M}_\alpha^\sigma \hat{\boldsymbol{\sigma}}_\alpha, \tag{15}$$

where $\boldsymbol{B}_{\alpha I} = \boldsymbol{B}_I(\boldsymbol{x}_\alpha)$ is the strain-displacement matrix (the smoothed strain-displacement matrix $\tilde{\boldsymbol{B}}_{\alpha I}$ is used in this study, refer to Appendix B.2), and $\boldsymbol{M}_\alpha^\sigma = \boldsymbol{M}^\sigma(\boldsymbol{x}_\alpha)$. As a result, the matrix equations of (13) result in:

$$\sum_{J=1}^{N} \left( \sum_{\alpha=1}^{m} V_\alpha \boldsymbol{B}_{\alpha I}^T \boldsymbol{M}_\alpha^\varepsilon \boldsymbol{B}_{\alpha J} \right) \boldsymbol{d}_J = \sum_{\alpha=1}^{m} V_\alpha \boldsymbol{B}_{\alpha I}^T \boldsymbol{M}_\alpha^\varepsilon \hat{\boldsymbol{\varepsilon}}_\alpha, \; I = 1, ..., N, \tag{16a}$$

$$\sum_{\alpha=1}^{m} V_\alpha \boldsymbol{B}_{\alpha I}^T \boldsymbol{\sigma}_\alpha = \boldsymbol{f}_I, \; I = 1, ..., N, \tag{16b}$$

$$\boldsymbol{M}_\alpha^\sigma \boldsymbol{\sigma}_\alpha - \sum_{I=1}^{N} \boldsymbol{B}_{\alpha I} \boldsymbol{\Lambda}_I = \boldsymbol{M}_\alpha^\sigma \hat{\boldsymbol{\sigma}}_\alpha, \; \alpha = 1, ..., m. \tag{16c}$$

where $\{V_\alpha\}_{\alpha=1}^m$ are the quadrature weights as defined in (7), and $\{\boldsymbol{f}_I\}_{I=1}^N$ are the nodal force vectors associated with the employed RK approximation of body force $\boldsymbol{b}$ and surface traction $\bar{\boldsymbol{t}}$. It can be seen that $\{\boldsymbol{d}_I\}_{I=1}^N$ are solved from (16a) directly, and $\{\boldsymbol{\Lambda}_I\}_{I=1}^N$ represent the displacement adjustment related to the difference between the computational stress and the given stress data $\{\hat{\boldsymbol{\sigma}}_\alpha\}_{\alpha=1}^m$, as shown in Equation (16c). Plugging (16c) into (16b) it yields

$$\sum_{J=1}^{N}\left(\sum_{\alpha=1}^{m} V_\alpha \boldsymbol{B}_{\alpha I}^T \boldsymbol{M}_\alpha^{\sigma-1} \boldsymbol{B}_{\alpha J}\right)\boldsymbol{\Lambda}_J = \boldsymbol{f}_I - \sum_{\alpha=1}^{m} V_\alpha \boldsymbol{B}_{\alpha I}^T \hat{\boldsymbol{\sigma}}_\alpha,\ I = 1,...,N, \quad (17a)$$

where $\{\boldsymbol{\Lambda}_I\}_{I=1}^N$ can be solved readily. Then, the computational stress $\{\boldsymbol{\sigma}_\alpha\}_{\alpha=1}^m$ are obtained by the following equation

$$\boldsymbol{\sigma}_\alpha = \hat{\boldsymbol{\sigma}}_\alpha + \boldsymbol{M}_\alpha^{\sigma-1}\sum_{I=1}^{N}\boldsymbol{B}_{\alpha I}\boldsymbol{\Lambda}_I,\ \alpha = 1,...,m. \quad (17b)$$

In summary, Equations (16a), (17a) and (17b) constitute the global step of the data-driven solver. In each global step, the displacement vector $\{\boldsymbol{d}_I\}_{I=1}^N$ is obtained from strain data $\{\hat{\boldsymbol{\varepsilon}}_\alpha\}_{\alpha=1}^m$ by complying with compatibility, while the displacement adjustment $\{\boldsymbol{\Lambda}_I\}_{I=1}^N$ is driven by the force residuals between the external force and the internal force computed by the experimental stress data $\{\hat{\boldsymbol{\sigma}}_\alpha\}_{\alpha=1}^m$, as shown in (17a).

In this study, we propose to use a stabilized conforming nodal integration (SCNI) [59] for the integration of the weak form (13) due to its nodal representation nature of both state and field variables at nodal points. The brief summary about SCNI can be found in Appendix B. In this approach, the continuum domain is partitioned by a Voronoi diagram (see **Fig. 15**), and both the state variables $\{(\boldsymbol{\varepsilon}_\alpha, \boldsymbol{\sigma}_\alpha)\}_{\alpha=1}^N = \{(\boldsymbol{\varepsilon}(\boldsymbol{x}_\alpha), \boldsymbol{\sigma}(\boldsymbol{x}_\alpha))\}_{\alpha=1}^N$ and the nodal displacement vectors $\{\boldsymbol{u}_I\}_{I=1}^N = \{\boldsymbol{u}(\boldsymbol{x}_I)\}_{I=1}^N$ are computed at the set of nodes located at $\{\boldsymbol{x}_\alpha\}_{\alpha=1}^N$, i.e. $m = N$. This approach minimizes the number of integration points where the stress and strain experimental data are searched in the local step (5b), allowing an enhanced effectiveness in the learning solver. The introduction of a smooth reproducing kernel (RK) shape function in the displacement approximation in (14a), such as the employment of a cubic B-spline in the RK approximation in Equations (B.1) - (B.7) in Appendix B, yields a $C^1$ continuous strain-displacement matrix $\boldsymbol{B}_I(\boldsymbol{x}_\alpha)$, and consequently a smooth tangent matrices in the displacement adjustment and stress update equations in (17a) and (17b), respectively. This smooth solution space of the physical manifold is made consistent with the continuous and convex solution space of the regularized LCDD learning solver to be introduced in Section 3.

### 2.3.2. *Local step of data-driven solver*

In this approach, the experimental data $\hat{\boldsymbol{s}} = (\hat{\boldsymbol{\varepsilon}}, \hat{\boldsymbol{\sigma}})$ as used for the global step solution (see (10) or (13)) is searched by the local step in (5b) during each local-global iteration. Considering the functional $\mathcal{H}$ as shown in (7), the local step of (5b) can be decomposed into $m$ local minimization problems: find *the optimal local data* $\hat{\boldsymbol{s}}_\alpha^* = (\hat{\boldsymbol{\varepsilon}}_\alpha^*, \hat{\boldsymbol{\sigma}}_\alpha^*)$, such that the distance to a given local state $\boldsymbol{s}_\alpha = (\boldsymbol{\varepsilon}_\alpha, \boldsymbol{\sigma}_\alpha)$ is minimized, i.e.

$$\hat{\boldsymbol{s}}_\alpha^* = \arg\min_{\hat{\boldsymbol{s}}_\alpha \in \mathbb{E}} d^2(\boldsymbol{s}_\alpha, \hat{\boldsymbol{s}}_\alpha) = \arg\min_{\hat{\boldsymbol{s}}_\alpha \in \mathbb{E}} d^{\varepsilon 2}(\boldsymbol{\varepsilon}_\alpha, \hat{\boldsymbol{\varepsilon}}_\alpha) + d^{\sigma 2}(\boldsymbol{\sigma}_\alpha, \hat{\boldsymbol{\sigma}}_\alpha), \tag{18}$$

for $\alpha = 1,...,m$.

### 2.3.3. *Standard data-driven solver*

The procedures for solving the data-driven problem are summarized below. Given a set of local data solutions $\{\hat{\boldsymbol{s}}_\alpha^{(\nu)}\}_{\alpha=1}^m \in \mathbb{E}_{em}$ (or $\hat{\boldsymbol{s}}_\alpha^{(\nu)} \in \mathbb{E}, \alpha = 1,...,m$ for homogeneous material) of the $\nu$-th iteration, the following global step and local step are iterated until convergence:

1) *Global Step*. Input: $\{\hat{\boldsymbol{s}}_\alpha^{(\nu)}\}_{\alpha=1}^m$ $\to$ Output: $\{\boldsymbol{s}_\alpha^{(\nu)}\}_{\alpha=1}^m$

   1.1 Solve Equations (16a) for $\{\boldsymbol{d}_I^{(\nu)}\}_{I=1}^N$ and (17a) for $\{\boldsymbol{\Lambda}_I^{(\nu)}\}_{I=1}^N$.

   1.2 Update computational states for $\{\boldsymbol{s}_\alpha^{(\nu)}\}_{\alpha=1}^m = \{(\boldsymbol{\varepsilon}_\alpha^{(\nu)}, \boldsymbol{\sigma}_\alpha^{(\nu)})\}_{\alpha=1}^m$ via

   $$\boldsymbol{\varepsilon}_\alpha^{(\nu)} = \sum_{I=1}^N \boldsymbol{B}_{\alpha I} \boldsymbol{d}_I^{(\nu)} \text{ and } \boldsymbol{\sigma}_\alpha^{(\nu)} = \hat{\boldsymbol{\sigma}}_\alpha^{(\nu)} + \boldsymbol{M}_\alpha^{\sigma -1} \sum_{I=1}^N \boldsymbol{B}_{\alpha I} \boldsymbol{\Lambda}_I^{(\nu)} \text{ in (17b)}$$

2) *Local Step*. Input: $\{\boldsymbol{s}_\alpha^{(\nu)}\}_{\alpha=1}^m \to$ Output: $\{\hat{\boldsymbol{s}}_\alpha^{(\nu+1)}\}_{\alpha=1}^m$

   for $\alpha = 1,...,m$, solve Equation (18) for $\hat{\boldsymbol{s}}_\alpha^{(\nu+1)}$.

**Remark 2.2**. It has been observed that distance minimizing data-driven (DMDD) computing solver [41] with distance measure in (7) - (9) is sensitive to data noise and outliers [42,54] because the local minimization stage (18) only searches for the nearest data point from the given experimental data set regardless of any latent data structure. The data-driven solution could be strongly influenced by the outliers locating near to the physical manifold $\mathcal{C}$ but do not conform to the hidden material data pattern (or the latent statistical model) of $\mathbb{E}$. Without the knowledge of the underlying data manifold, it requires a large amount of data to achieve sufficiently accurate predictions which is costly [20,31].

## 3. Local convexity-preserving data-driven approach

In this section, we introduce the LCDD approach, which introduces the locally convex reconstruction technique inspired by manifold learning strategies [21,62,63]. Our aim in this work is to develop a computationally feasible data-driven predictive modeling framework to enhance accuracy and robustness against noise and outliers in the experimental data set by constructing the local manifold with the desired smoothness and convexity.

For ease of exposition, we define a weighted vector norm "$\|\cdot\|_M$" based on (8) as follows

$$\| \boldsymbol{s}_\alpha \|_M^2 = \boldsymbol{s}_\alpha^T \bar{\boldsymbol{M}}_\alpha \boldsymbol{s}_\alpha \equiv \| \bar{\boldsymbol{M}}_\alpha^{1/2} \boldsymbol{s}_\alpha \|^2, \tag{19}$$

where $\boldsymbol{s}_\alpha = [\boldsymbol{\varepsilon}_\alpha^T \; \boldsymbol{\sigma}_\alpha^T]^T \in \mathbb{R}^{2q}$, $\bar{\boldsymbol{M}}_\alpha = diag([\boldsymbol{M}_\alpha^\varepsilon, \boldsymbol{M}_\alpha^\sigma])$, and $\bar{\boldsymbol{M}}_\alpha^{1/2}$ can be determined by the singular value decomposition of $\bar{\boldsymbol{M}}_\alpha$. For a given $\boldsymbol{s}_\alpha$, the local step (18) is rewritten as

$$\hat{\boldsymbol{s}}_\alpha^* = \arg\min_{\hat{\boldsymbol{s}}_\alpha \in \mathbb{E}} \| \boldsymbol{s}_\alpha - \hat{\boldsymbol{s}}_\alpha \|_M^2, \tag{20}$$

for $\alpha = 1, ..., m$.

### *3.1. Locally convex construction*

It has been shown [21,62,64–66] that naturally occurring data usually reside on a lower dimensional submanifold which is embedded in the high-dimensional ambient space, as shown in **Fig. 2**. In this study, inspired by locally linear embedding (LLE) approach [62], we assume there exists an underlying manifold of low dimensionality corresponding to the raw experimental data set, i.e. $\mathbb{E} = \{\hat{\boldsymbol{s}}^i, i = 1, ..., p\}$ where $p$ is the number of data points, that is locally linear and smooth varying. Therefore, a data point $\hat{\boldsymbol{s}}^i \in \mathbb{E}$ can be linearly reconstructed from its neighbors in the data set, i.e.

$$\hat{\boldsymbol{s}}^i \approx \hat{\boldsymbol{s}}_{recon}^i = \sum_{j \in \mathcal{N}_k(\hat{\boldsymbol{s}}^i)} w_{ij} \hat{\boldsymbol{s}}^j, \tag{21}$$

where $\hat{\boldsymbol{s}}_{recon}^i$ is the reconstruction of $\hat{\boldsymbol{s}}^i$, $\mathcal{N}_k(\hat{\boldsymbol{s}}^i)$ is the set of the $k$ nearest neighbor ($k$-NN) data points to $\hat{\boldsymbol{s}}^i$ in $\mathbb{E}$, and $w_{ij}$ are the unknown coefficients. In LLE, the optimal reconstruction weights $w_{ij}^*$ can be obtained by solving the following problem:

$$
\begin{aligned}
&\{w_{ij}^*\}_{i,j=1,\ldots,p} = \arg\min \sum_{i=1}^{p} \| \hat{\mathbf{s}}^i - \sum_{j=1, j\neq i}^{p} w_{ij} \hat{\mathbf{s}}^j \|^2 \\
&\text{subject to: } \sum_{j=1}^{p} w_{ij} = 1, \ i = 1, \ldots, p \\
&\qquad\qquad w_{ij} = 0 \ \text{ if } \ j \notin \mathcal{N}_k(\hat{\mathbf{s}}^i)
\end{aligned}
\tag{22}
$$

Note that $w_{ij} = 0$ when $i = j$. The data reconstruction procedures in (21) and (22) provide the projection of $\hat{\mathbf{s}}^i$, i.e. $\hat{\mathbf{s}}^i_{recon}$, onto the subspace spanned by $\{\hat{\mathbf{s}}^j\}_{j \in \mathcal{N}_k(\hat{\mathbf{s}}^i)}$ with respect to the norm.

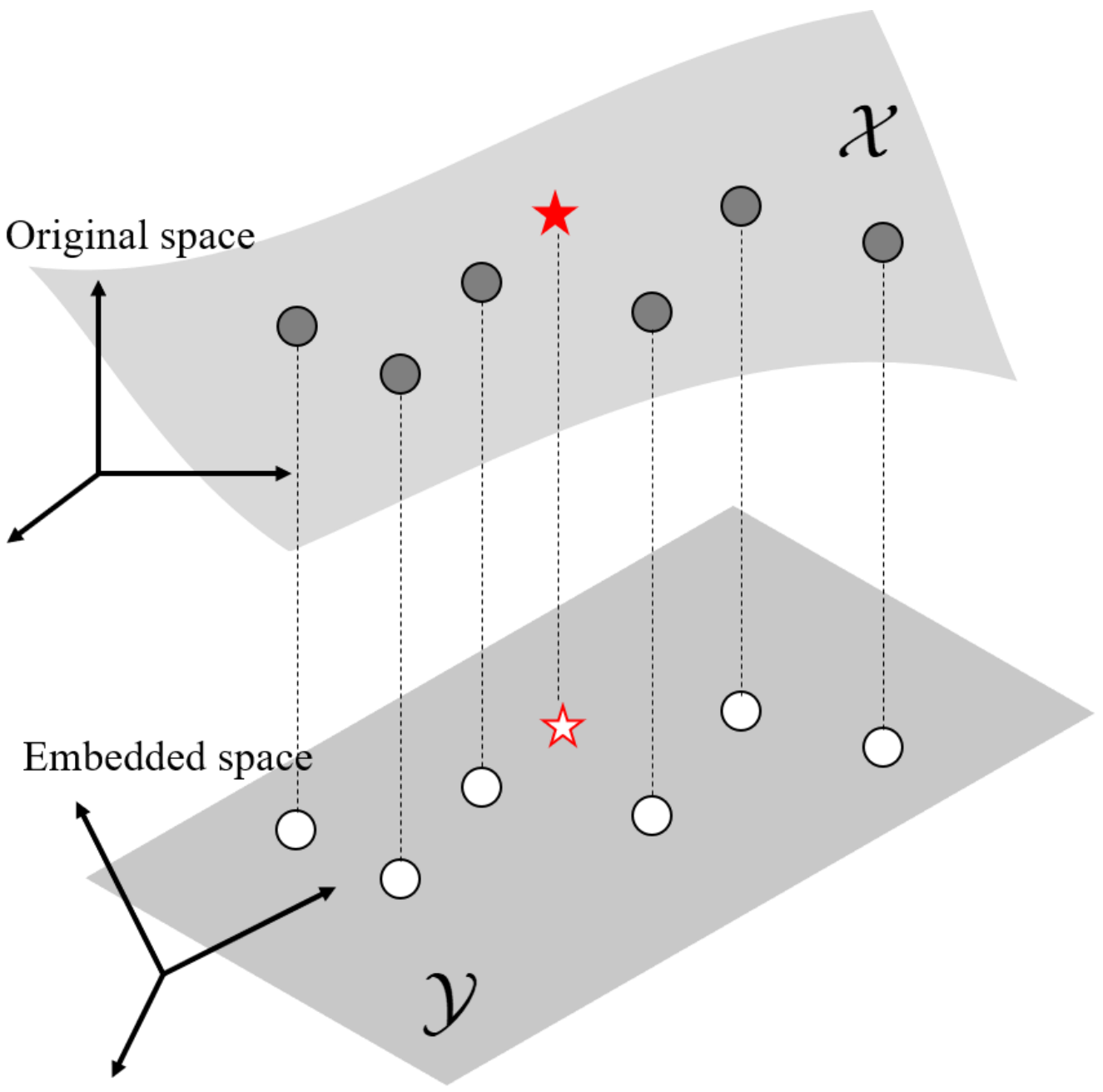

**Fig. 2.** Schematic of a manifold embedded in the original space and the associated low-dimensional embedding, where the training samples and the new sample are denoted by gray circles and red star, respectively.

Different from the standard LLE, the search for solution data point in data-driven computing is constrained by the physical manifold associated with (1) and (2). But considering that the data-driven algorithm in (5) performs a fixed point iteration on the experimental data points that are closest to the physical manifold, the locally linear reconstruction remains suitable for this scenario. In this sense, it is similar to the out-of-sample extension problem [63,67,68] where it is desirable to add new projected data points (new samples) to a previously learnt low-dimensional embedding, as shown in **Fig. 2**. In addition, from physical perspective, the data-driven solution constrained by the physics laws need to be close enough to the graph of the experimental data with underlying constitutive data structure. Thus, we need to prevent the reconstructed data point by (21) from projecting to a point that is far away from the underlying material data structure on the embedded subspace. To this end, we propose a local manifold learning algorithm to reconstruct the given local state on the locally convex manifold of the experimental data set.

Given a local state $\boldsymbol{s}_\alpha$, the most representative $k$ nearest neighbor ($k$-NN) points in $\mathbb{E}$ are first identified using the same metric induced by the given norm "$\|\cdot\|_M$", and collected as $\{\hat{\boldsymbol{s}}^i\}_{i\in\mathcal{N}_k(\boldsymbol{s}_\alpha)} \subset \mathbb{E}$, in which the indices for the nearest neighbors of $\boldsymbol{s}_\alpha$ are stored in a set $\mathcal{N}_k(\boldsymbol{s}_\alpha)$. Then we project the local state onto the convex hull of $\{\hat{\boldsymbol{s}}^i\}_{i\in\mathcal{N}_k(\boldsymbol{s}_\alpha)}$ associated to $\boldsymbol{s}_\alpha$, which is defined as:

$$\mathcal{E}(\boldsymbol{s}_\alpha) = \text{Conv}(\{\hat{\boldsymbol{s}}^i\}_{i\in\mathcal{N}_k(\boldsymbol{s}_\alpha)}) = \left\{ \sum_{i\in\mathcal{N}_k(\boldsymbol{s}_\alpha)} w_i \hat{\boldsymbol{s}}^i \,\middle|\, \sum_{i\in\mathcal{N}_k(\boldsymbol{s}_\alpha)} w_i = 1, \text{ and } w_i \geq 0, \forall i \in \mathcal{N}_k(\boldsymbol{s}_\alpha) \right\}, \quad (23)$$

or concisely denoted as $\mathcal{E}_\alpha$. Accordingly, the optimal reconstruction coefficients are given by solving the following minimization problem:

$$\begin{aligned} \boldsymbol{w}_\alpha^* &= \arg\min_{\boldsymbol{w}\in\mathbb{R}^k} \| \boldsymbol{s}_\alpha - \sum_{i\in\mathcal{N}_k(\boldsymbol{s}_\alpha)} w_i \hat{\boldsymbol{s}}^i \|_M^2 \\ \text{subject to: } & \sum_{i\in\mathcal{N}_k(\boldsymbol{s}_\alpha)} w_i = 1, \\ & w_i \geq 0, \ \forall i \in \mathcal{N}_k(\boldsymbol{s}_\alpha), \end{aligned} \quad (24a)$$

where $\boldsymbol{w} \in \mathbb{R}^k$ denotes the vector consisting of the weights $\{w_i\}_{i\in\mathcal{N}_k(\boldsymbol{s}_\alpha)}$ corresponding to the $k$ selected neighbor points, and $\boldsymbol{w}_\alpha^*$ with the subscript $\alpha$ denotes the optimal weights corresponding to the given local state $\boldsymbol{s}_\alpha$. The reconstruction $\hat{\boldsymbol{s}}_\alpha^*$ can be retrieved by using the linear combination of $\{\hat{\boldsymbol{s}}^i\}_{i\in\mathcal{N}_k(\boldsymbol{s}_\alpha)}$ with the computed weight vector $\boldsymbol{w}_\alpha^*$ as follows

$$\hat{\boldsymbol{s}}_\alpha^* = \sum_{j\in\mathcal{N}_k(\boldsymbol{s}_\alpha)} w_j^* \hat{\boldsymbol{s}}^j = \hat{\boldsymbol{S}}_\alpha \boldsymbol{w}_\alpha^*, \quad (24b)$$

where $\hat{\boldsymbol{S}}_\alpha \in \mathbb{R}^{2q\times k}$ is the matrix composed of the *k*-NN data points $\{\hat{\boldsymbol{s}}^i\}_{i\in\mathcal{N}_k(\boldsymbol{s}_\alpha)}$. This approach in (24) is called *locally convex construction*. Compared to the equation (22), the main differences in (24a) are: 1. a new data point $\boldsymbol{s}_\alpha$ obtained from the physical solver, instead of other points in the experimental data set, is used for local construction; 2. a weighted vector norm " $\|\cdot\|_M$ " representing energy is adopted for distance measure.

Based on the idea of locally convex construction, the local step of data-driven computing in (20) is modified as: Given the data-set neighbors $\{\hat{\boldsymbol{s}}^i\}_{i\in\mathcal{N}_k(\boldsymbol{s}_\alpha)} \subset \mathbb{E}$ for $\boldsymbol{s}_\alpha$, solve $\hat{\boldsymbol{s}}_\alpha^*$ such that

$$\hat{\boldsymbol{s}}_\alpha^* = \arg\min_{\hat{\boldsymbol{s}}_\alpha \in \mathcal{E}_\alpha} \| \boldsymbol{s}_\alpha - \hat{\boldsymbol{s}}_\alpha \|_M^2, \tag{25}$$

for $\alpha = 1,...,m$. By comparing (20) and (25), we can observe that the space $\mathbb{E}$ used in the standard data-driven scheme [41,54] is now replaced by the associated convex hull $\mathcal{E}_\alpha$ that is locally reconstructed around the input $\boldsymbol{s}_\alpha$ by learning techniques, allowing to capture the local material manifold. Consequently, the reconstruction data (i.e., the optimal local data) $\hat{\boldsymbol{s}}_\alpha^*$ is sought from the set $\mathcal{E}_\alpha$ with convexity and smoothness. With the definition in (23), the solution of the minimization problem (25) is obtained by solving (24).

**Remark 3.1**. Equation (24a) is a *constrained regression* or *constrained least-squares* problem under a *invariance* constraint and a *non-negative* constraint. The invariance constraint imposes the partition of unity on the weight array $\boldsymbol{w}$, i.e. $\boldsymbol{1}^T\boldsymbol{w} = 1$, where $\boldsymbol{1} = [1,1,...,1]^T \in \mathbb{R}^k$. It ensures the invariance of the reconstruction weights $\boldsymbol{w}_\alpha^*$ to rotations, rescaling, and translations of the same *k*-NN data points, and thus, the weights characterize geometric properties independent of a particular frame of reference [62,63]. It also guarantees the linear approximation property such that $\hat{\boldsymbol{s}}_\alpha^*$ is in the subspace $\mathrm{span}(\{\hat{\boldsymbol{s}}^i\}_{i\in N_k(\boldsymbol{s}_\alpha)})$. When we further consider the non-negative constraint, the approximation $\hat{\boldsymbol{s}}_\alpha^*$ is restricted to the convex hull $\mathcal{E}(\boldsymbol{s}_\alpha)$ (see **Fig. 3**). The imposed convexity and locality yields enhanced robustness of linear regressions to outliers [63,69], and reduces numerical instability across different clusters of neighbor points during data-driven iterations. Moreover, it is well known that the non-negative constraint naturally imposes sparseness on the coefficient solution $\boldsymbol{w}_\alpha^*$. Lastly, by specifying the number of *k*-NN points, it provides an opportunity to incorporate a priori knowledge about the experimental data structure and therefore, enhance the robustness of data learning [63].

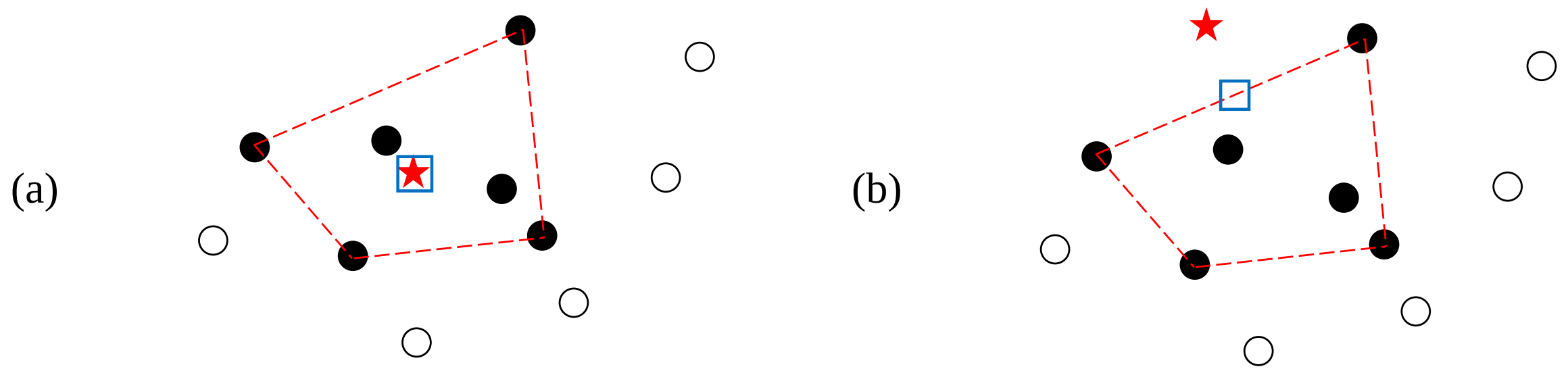

**Fig. 3.** Sketch of the projection $\hat{\mathbf{s}}_\alpha^*$ (the blue square) on a convex hull $\mathcal{E}_\alpha$ (the region is depicted by red dashed lines) of *k*-NN points (the solid circles in black) when a local state $\boldsymbol{s}_\alpha$ (the red star) locates (a) inside and (b) outside $\mathcal{E}_\alpha$. Neighbor points of $k = 6$ are used for demonstration.

Essentially, the modified local step of data-driven problem in (25) can be interpreted as a process of seeking the data approximation based on the previously learnt low-dimensional manifold $\mathcal{E}_\alpha$ associated with the given local state. From a geometrical point of view, it searches the projection (i.e. the optimal local material data $\hat{\mathbf{s}}_\alpha^* \in \mathcal{E}_\alpha$) in the associated convex set $\mathcal{E}_\alpha$. If $\boldsymbol{s}_\alpha$ locates inside $\mathcal{E}_\alpha$, the projection is represented by $\boldsymbol{s}_\alpha$ itself (**Fig. 3**a). Otherwise, the local state is optimally projected on the convex hull $\mathcal{E}_\alpha$ and the projection point is considered as the best representative on the constitutive manifold (**Fig. 3**b).

### *3.2. Solving non-negative least squares*

In this section, a computationally feasible algorithm is developed to solve the local step minimization problem in (24a) by relating it to the *non-negative least squares* (NNLS) problem that has been well established. The NNLS problem is reviewed in Append A.

However, to solve the minimization problem (24a) under the NNLS framework, see (A.1) , the partition of unity constraint in (24a) needs to be properly handled. In this end, we propose to employ the quadratic penalty method [70] to penalize the residuals of the partition of unity constraint in the auxiliary objective, and the modified minimization problem becomes

$$\begin{aligned} &\boldsymbol{w}_\alpha^* = \arg\min_{\boldsymbol{w}\in\mathbb{R}^k} \| \boldsymbol{s}_\alpha - \hat{\boldsymbol{S}}_\alpha \boldsymbol{w} \|_M^2 + \xi (\boldsymbol{1}^T \boldsymbol{w} - 1)^2, \\ &\text{subject to: } w_i \geq 0,\ i = 1,...,k, \end{aligned} \tag{26}$$

where $\xi > 0$ is a regularized coefficient to impose the associated constraint.

Note that to conform with the Euclidean metric used in the standard NNLS solver (A.1), the local state $\boldsymbol{s}_\alpha$ can be easily rescaled to $\boldsymbol{z}_\alpha = \bar{\boldsymbol{M}}_\alpha^{1/2} \boldsymbol{s}_\alpha$ (similarly, $\hat{\boldsymbol{Z}}_\alpha = \bar{\boldsymbol{M}}_\alpha^{1/2} \hat{\boldsymbol{S}}_\alpha$) by using the relation given in (19). As a result, the minimization problem (26) can be recast into a NNLS form as shown by augmenting the vector $\boldsymbol{s}_\alpha$ and the matrix $\hat{\boldsymbol{Z}}_\alpha$ with additional components as follows

$$\begin{aligned} &\boldsymbol{w}_\alpha^* = \arg\min_{\boldsymbol{w}\in\mathbb{R}^k} \| \boldsymbol{z}_\alpha^{\text{aug}} - \hat{\boldsymbol{Z}}_\alpha^{\text{aug}} \boldsymbol{w} \|^2, \\ &\text{subject to: } w_i \geq 0,\ i = 1,...,k, \end{aligned} \tag{27}$$

where

$$\hat{\boldsymbol{Z}}_\alpha^{\text{aug}} := \begin{bmatrix} \hat{\boldsymbol{Z}}_\alpha \\ \sqrt{\xi}\,\boldsymbol{1}^T \end{bmatrix} \in \mathbb{R}^{(2q+1)\times k},\ \boldsymbol{z}_\alpha^{\text{aug}} := \begin{bmatrix} \boldsymbol{z}_\alpha \\ \sqrt{\xi} \end{bmatrix} \in \mathbb{R}^{2q+1}. \tag{28}$$

To properly impose the penalty term, we set $\xi = \bar{\xi}\,\text{tr}(\hat{\boldsymbol{Z}}_\alpha^T \hat{\boldsymbol{Z}}_\alpha)/k$, where $\bar{\xi}$ is a large parameter (usually set as $10^4 - 10^6$). With the weight solution $\boldsymbol{w}_\alpha^*$ solved by NNLS algorithm, the reconstruction $\hat{\boldsymbol{s}}_\alpha^*$ can be obtained via (24b).

It is possible that constrained least squares in (27) could suffer from numerical instability due to rank deficiency when the number of neighbors is larger than the rank of the neighborhood, i.e. $k > \text{rank}(\hat{\boldsymbol{Z}}_\alpha^{\text{aug}})$. As has been well studied in machine learning field [5], a further regularization can be introduced to the NNLS problem. In this study, a commonly used ridge regression [71], or called Tikhonov regularization, is applied to address the ill-posed issues, and the NNLS problem (27) is modified as

$$\begin{aligned} &\boldsymbol{w}_\alpha^* = \arg\min_{\boldsymbol{w}\in\mathbb{R}^k} \| \hat{\boldsymbol{Z}}_\alpha^{\text{aug}} \boldsymbol{w} - \boldsymbol{z}_\alpha^{\text{aug}} \|^2 + \mu \| \boldsymbol{w} \|^2, \\ &\text{subject to: } w_i \geq 0,\ i = 1,...,k, \end{aligned} \tag{29}$$

where the regularized coefficient is

$$\mu = \bar{\mu}\,\text{tr}(\hat{\boldsymbol{Z}}_\alpha^T \hat{\boldsymbol{Z}}_\alpha)/k, \tag{30}$$

Here $\bar{\mu}$ is a small constant (set as $10^{-4}$ by default) such that the regularization has minor effect on the solution $\boldsymbol{w}_\alpha^*$ and the reconstruction $\hat{\boldsymbol{s}}_\alpha^*$. It is also shown that [5] this regularization imposes certain smoothness on the solution and guarantees a unique solution.

**Remark 3.2.** As discussed in [63], the size range of *k*-NN points depends on various features of the data, such as the manifold geometry and the sampling density. In principle, *k* should be greater than the underlying manifold dimensionality of the material data set $\mathbb{E}$ in order to explore the data structural and prevent overwhelming influence from outliers/noise. Meanwhile, the resultant neighborhoods should be localized enough to ensure the validity of locally linear approximation.

### *3.3. Local convexity-preserving data-driven solver*

A simple algorithm for the proposed LCDD solver is shown as follows: Given a convergence tolerance TOL and the material database $\mathbb{E}$, then

1. Initialize $\hat{\boldsymbol{s}}_\alpha^{*(0)} = [\hat{\boldsymbol{\varepsilon}}_\alpha^{*(0)T} \; \hat{\boldsymbol{\sigma}}_\alpha^{*(0)T}]^T$, $\alpha = 1, ..., m$ randomly, and $v = 0$.
2. While $\max_{\alpha=1,...,m} \| \hat{\boldsymbol{s}}_\alpha^{*(v)} - \hat{\boldsymbol{s}}_\alpha^{*(v-1)} \|_M > TOL$
   a. Solve equations (16), and output $\{\boldsymbol{s}_\alpha^{(v)}\}_{\alpha=1}^m$
   b. Construct *k*-NN neighborhood $\mathcal{N}_k(\boldsymbol{s}_\alpha^{(v)})$ and $\hat{\boldsymbol{S}}_\alpha$ for each local state $\boldsymbol{s}_\alpha^{(v)}$.
   c. Solve NNLS (27) (or (29)) by **Algorithm 1**, and use $\boldsymbol{w}_\alpha^*$ to output $\hat{\boldsymbol{s}}_\alpha^{*(v+1)}$ via (24b).
   d. Update: $v \leftarrow v + 1$
3. Solution is $\boldsymbol{s}_\alpha = [\boldsymbol{\varepsilon}_\alpha^T \; \boldsymbol{\sigma}_\alpha^T]^T \leftarrow \boldsymbol{s}_\alpha^{(v)} = [\boldsymbol{\varepsilon}_\alpha^{(v)T} \; \boldsymbol{\sigma}_\alpha^{(v)T}]^T$, $\alpha = 1, ..., m$.

It has been shown that the Lawson-Hanson method [72,73] (**Algorithm 1** in Appendix A) used for solving NNLS converges in a finite number of iterations less than the size of the output coefficient vector, which is the size of *k*-NN in LCDD. In addition, considering the small size of the local matrix $\hat{\boldsymbol{S}}_\alpha \in \mathbb{R}^{2q \times k}$ , $k, q \ll \min(N, m)$ , where $N$ and $m$ are the numbers of discretization nodes and integration points, respectively, the additional computational cost in solving the NNLS problem in (27) or (29) is negligible compared to solving the linear system (16 ).

## 4. Numerical Examples: Truss Structures

In this section, the accuracy, convergence and robustness properties of the proposed LCDD approach are examined. Synthetic data sets are employed to verify the performance of the proposed method. The standard DMDD approach without considering any local data structure is also provided for comparison. For simplicity, we consider homogeneous material in the following numerical examples in Section 4 and Section 5. That is, the same material database $\mathbb{E} = \{\hat{\boldsymbol{s}}^i = (\hat{\boldsymbol{\varepsilon}}^i, \hat{\boldsymbol{\sigma}}^i), i = 1, ..., p\}$, where $p$ is the number of data points, is introduced for all material points.

In this section, the material behavior of the $\alpha$-th bar member is characterized by a simple uniaxial strain $\varepsilon_\alpha$ and uniaxial stress $\sigma_\alpha$ relationship. As such, the local state vector is defined as $\boldsymbol{s}_\alpha = [\varepsilon_\alpha \ \ \sigma_\alpha]^T \in \mathbb{R}^2$, and the associated norm to measure the distances of local states is given as

$$\| \boldsymbol{s}_\alpha \|_M = \left( \frac{1}{2} M \varepsilon_\alpha^2 + \frac{1}{2} M^{-1} \sigma_\alpha^2 \right)^{1/2}, \tag{31}$$

where $M$ is a positive constant analogous to the Young's modulus of the reference material.

In the following examples, $k$ is the number of the $k$-NN used in the local step of the LCDD solver and $\bar{\mu}$ is the regularization coefficient in (30) with a default value $10^{-4}$.

### 4.1. Example I: One-dimensional truss

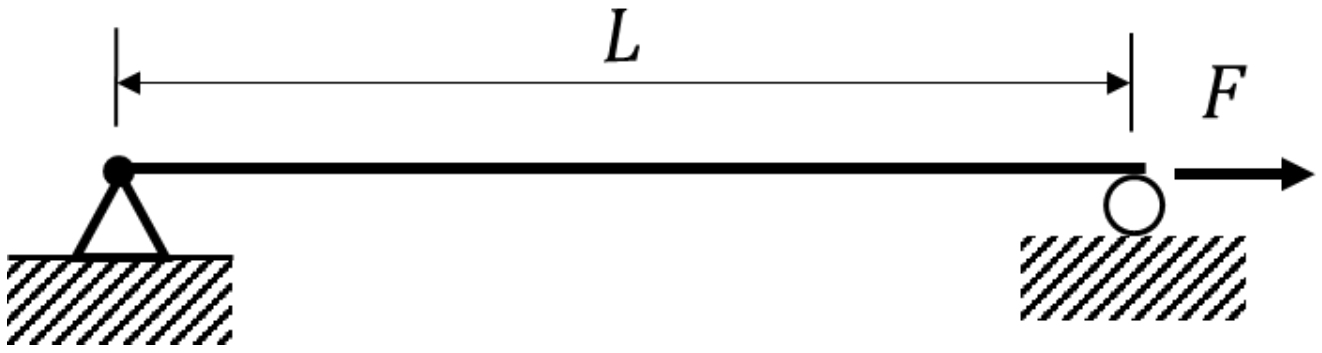

**Fig. 4.** One-bar truss structure with the cross-section area $A = 200 \text{ cm}^2$ subjected to a uniaxial load $F = 10 \text{ kN}$.

This example examines data-driven computing in a single truss member ($m = 1$) when dealing with irregular material data that exhibits noise and outliers. A truss member with the cross-section area $A = 200 \text{ cm}^2$ is subjected to an axial load of $F = 10 \text{ kN}$ as shown in **Fig. 4**.

*4.1.1. Data set with random noise*

To examine the performance of LCDD when dealing with noisy material data, two material databases (denoted by the circle points) with different levels of Gaussian random noise (i.e. $\chi = 0.05$ and $\chi = 0.15$ in (32)) are considered, as shown in **Fig. 5** and **Fig. 6**. The data sets are given by a set of noiseless strain and stress points, i.e. $\{\bar{\varepsilon}^i, \bar{\sigma}^i\}_{i=1}^p$, superimposed with a Gaussian noise as follows:

$$\begin{aligned} \hat{\varepsilon}^i &= \bar{\varepsilon}^i + \mathcal{N}(0, \chi\bar{\varepsilon}_{\max}),\ i = 1,\dots, p \\ \hat{\sigma}^i &= \bar{\sigma}^i + \mathcal{N}(0, \chi M\bar{\varepsilon}_{\max}),\ i = 1,\dots, p \end{aligned} \tag{32}$$

where the coefficient constant is $M = 100$ MPa, and $\chi$ is the given random level. In (32), the noiseless strain data $\{\bar{\varepsilon}^i\}_{i=1}^p$ is generated by a uniform distribution within a given strain range $[-\bar{\varepsilon}_{\max}, \bar{\varepsilon}_{\max}]$ with $\bar{\varepsilon}_{\max} = 0.01$, the noiseless stress data $\{\bar{\sigma}^i\}_{i=1}^p$ is defined as $\bar{\sigma}^i = M\bar{\varepsilon}^i$, and the Gaussian noise $\mathcal{N}(0, \chi\bar{\varepsilon}_{\max})$ and $\mathcal{N}(0, \chi M\bar{\varepsilon}_{\max})$ of strain and stress represent the normal distribution with zero mean and standard deviations of $\chi\bar{\varepsilon}_{\max}$ and $\chi M\bar{\varepsilon}_{\max}$, respectively.

The external force is incrementally loaded via 5 equal steps at the one-bar truss structure in **Fig. 4**, and thus 5 corresponding incremental data-driven solutions (depicted by the asterisk points) are shown in **Fig. 5** and **Fig. 6**. The associated optimal local data to the minimization problem in (25), i.e. $\hat{\mathbf{s}}_\alpha^* = [\hat{\varepsilon}_\alpha^* \ \hat{\sigma}_\alpha^*]^T$, at these 5 loading steps are denoted by the triangle points. As shown in **Fig. 5**, both the data-driven solutions of DMDD and LCDD can finally converge to a state point at the physical (equilibrium) manifold of $\sigma = 0.5$ Mpa and stay close to the given data set. Although the mild randomness is considered, DMDD yields a less desirable result compared to the result obtained by LCDD with the number of neighbor points $k = 6$. The data-driven solution from LCDD converges at the intersection of the physical manifold and the conjectural material graph, which is an ideal solution as discussed in Section 2. Note that by using the proposed locally convex reconstruction (see (23) or (24)), the local solver in (25) allows to attain the optimal local data from the reconstructed local convex hull instead of a direct search in the raw experimental points.

When the given material database presents stronger randomness, as shown in **Fig. 6**, DMDD performs poorly as it prematurely converges to a suboptimal solution far outside the data set, and its incremental solutions attain some local minimum points (**Fig. 6**a), implying susceptibility to data noise. In contrast, LCDD yields more desirable results (see **Fig. 6**b), where the incremental data-driven solutions move consistently to the intersection of the equilibrium manifold and the underlying material submanifold.

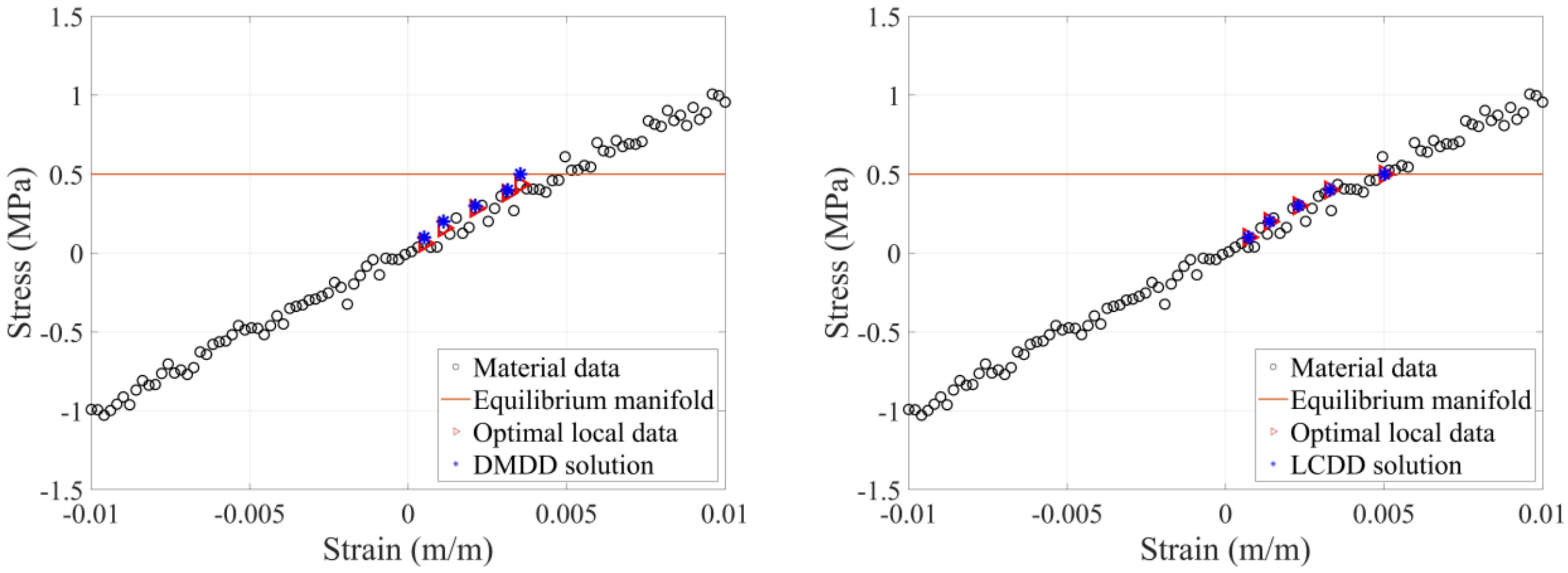

(a) (b)

**Fig. 5.** Comparison of the (a) DMDD and (b) LCDD solvers for the one-bar truss structure using a material database with mild Gaussian random noise $\chi = 0.05$. The database contains $p = 100$ stress-strain data points. The number of neighbor points used in LCDD is $k = 6$.

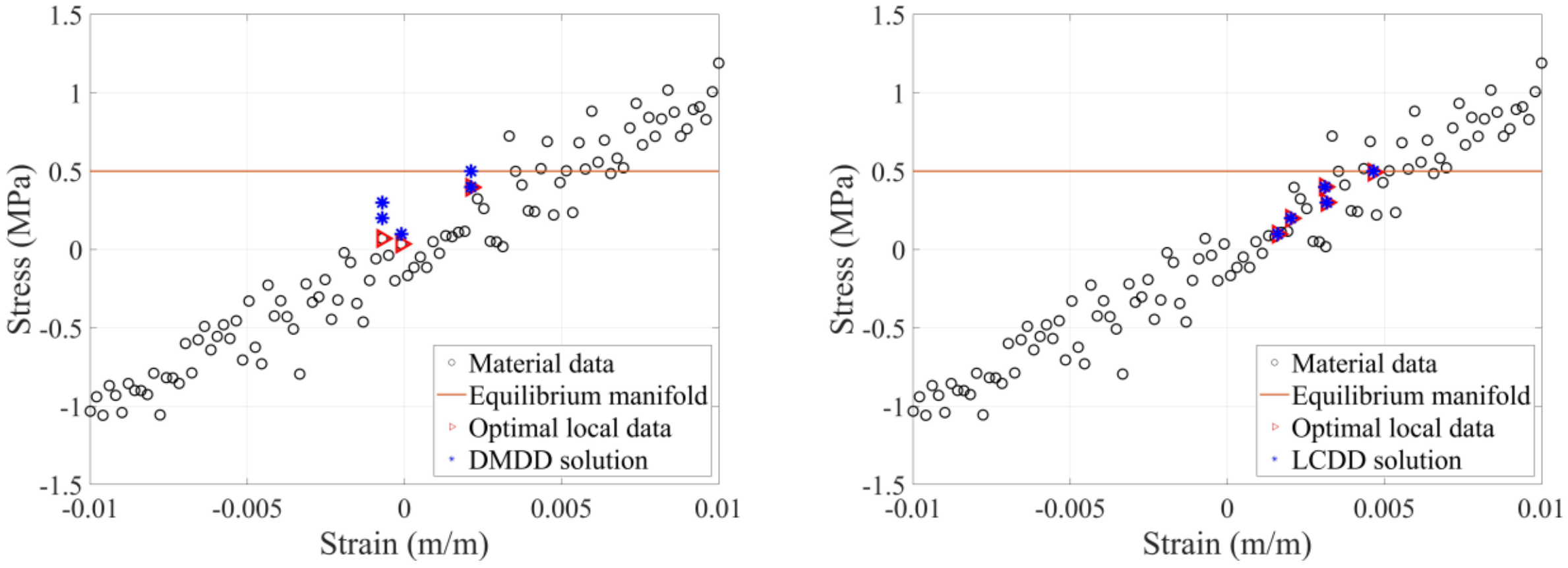

(a) (b)

**Fig. 6.** Comparison of the (a) DMDD and (b) LCDD solvers for the one-bar truss structure using a material database with strong Gaussian random noise $\chi = 0.15$. The database contains $p = 100$ stress-strain data points. The number of neighbor points used in LCDD is $k = 6$.

The comparison of the results in **Fig. 5** and **Fig. 6** suggests that LCDD is a robust scheme against large noise and local minimum wells, yielding similar pattern of convergence with material databases of different random levels, whereas the iterative solution path of DMDD is sensitive to the level and distribution of noise. It is also observed that the optimal data solutions $(\hat{\varepsilon}_\alpha^*, \hat{\sigma}_\alpha^*)$ obtained from LCDD usually coincide with the data-driven solutions $(\varepsilon_\alpha, \sigma_\alpha)$ even though no experiment data in $\mathbb{E}$ is exactly at those locations. This study indicates the advantage of LCDD in forming an implicit local material graph (via the convex hull) for searching the optimal data points. This unique feature allows LCDD to capture the local data structure, providing not only robustness against noise due to clustering analysis, but also the reproducibility to a locally linear manifold if the data is well sampled, which will be further discussed in the following examples.

### *4.1.2. Data set with outliers*

The robustness of the proposed LCDD solver is further demonstrated by consider a dataset with outliers. Let a material data set represent an underlying linear graph, i.e.

$F(\hat{\varepsilon}^i,\hat{\sigma}^i)=\hat{\sigma}^i-M\hat{\varepsilon}^i=0$ with an outlier positioned near to the linear manifold, as shown in **Fig. 7**. The data-driven solutions and the associated local projected data at each incremental step are plotted in **Fig. 7**. The results show that the DMDD solutions are misled by the presence of the outlier (**Fig. 7**a), whereas the LCDD solutions successfully converge to a reasonable location at the material manifold (**Fig. 7**b). In addition, to verify the effect of the locally convex construction parameters on the performance of the LCDD solver, two exemplary results using a different *k*-NN numbers ($k=12$) and a different regularization coefficient ($\bar{\mu}=10^{-2}$) are provided in **Fig. 7**c and **Fig. 7**d, respectively. As shown in **Fig. 7**c, while using more neighbor points in LCDD tends to increase the probability to involve the outlier for local manifold construction and causes the intermediate results during simulation more influenced by the outlier, it yields a final data-driven solution consistent to the one using $k=6$. This robustness is achieved from the clustering analysis based on the reconstruction of multiple data points that prevents the dominance of outliers. The effect of the regularization coefficient $\bar{\mu}$ on LCDD is also studied, see **Fig. 7**d. It shows that with larger $\bar{\mu}$ the optimal local data solution sought from the associated convex hull favors the region with higher data density because the reconstructed weights of the outliers have been penalized (refer to (29)). In this study, $\bar{\mu}=10^{-4}$ is adopted as the default setting. There are other regularization methods and robust penalty functions that allow further suppressing the influence of noise or outliers, e.g. the Huber penalty function. The interested readers are encouraged to consult the reference [5].

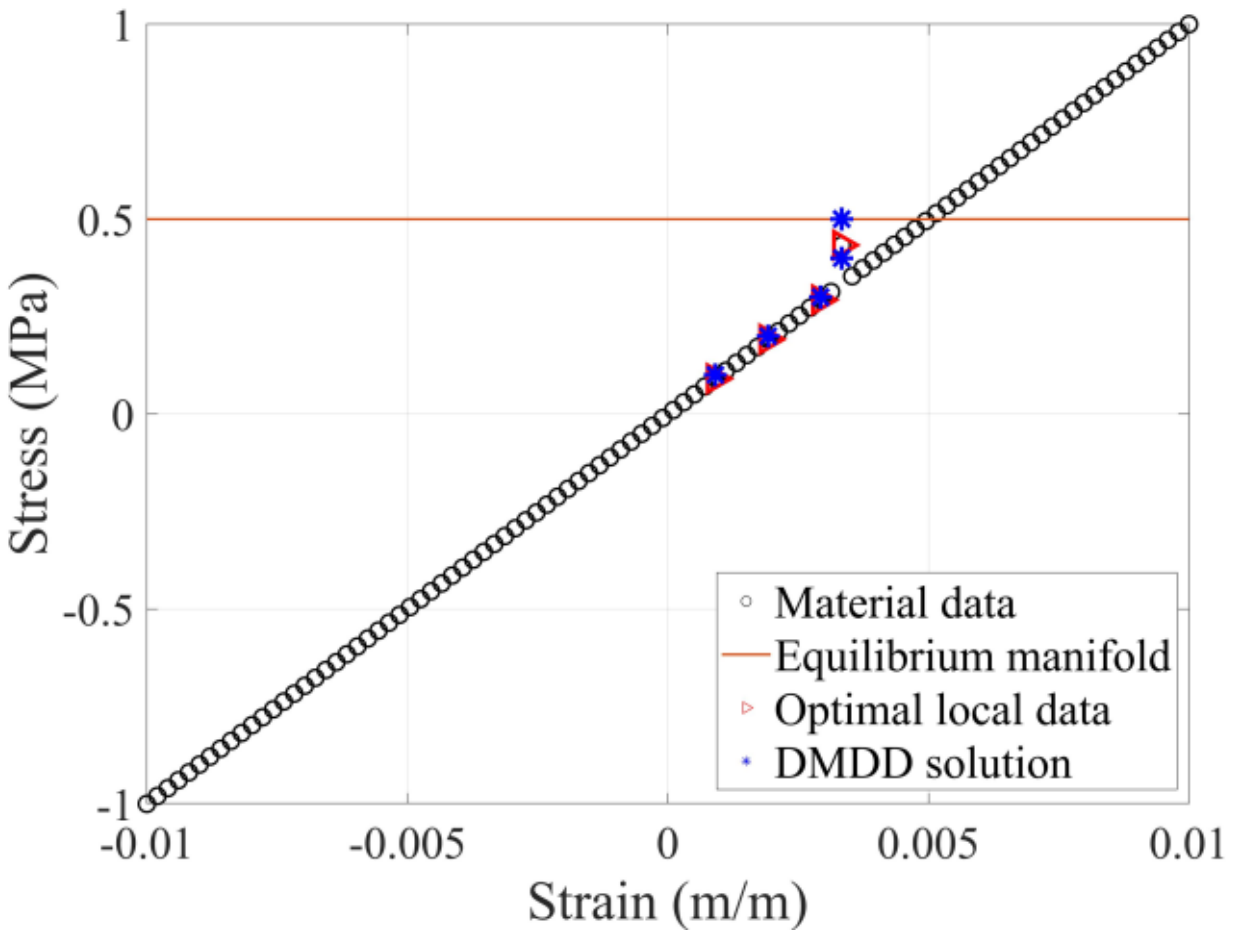

(a) DMDD

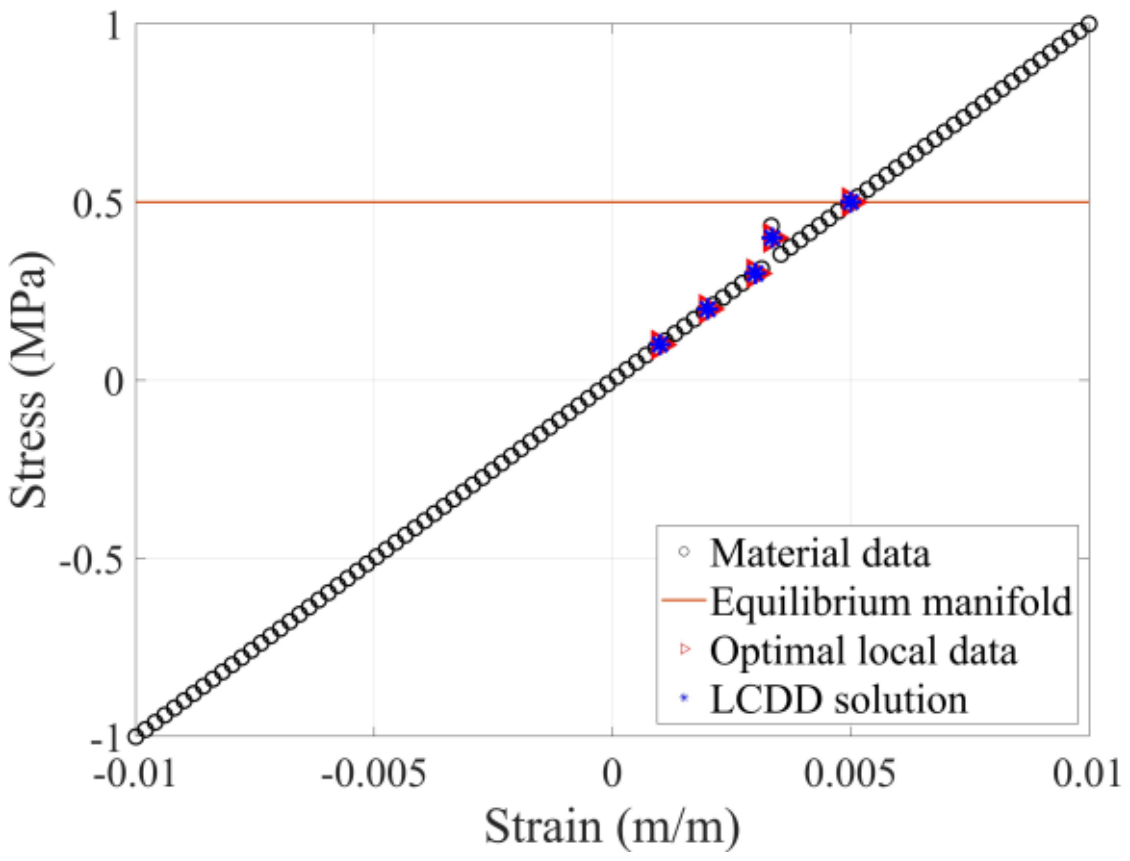

(b) LCDD: $k = 6$, $\bar{\mu} = 10^{-4}$

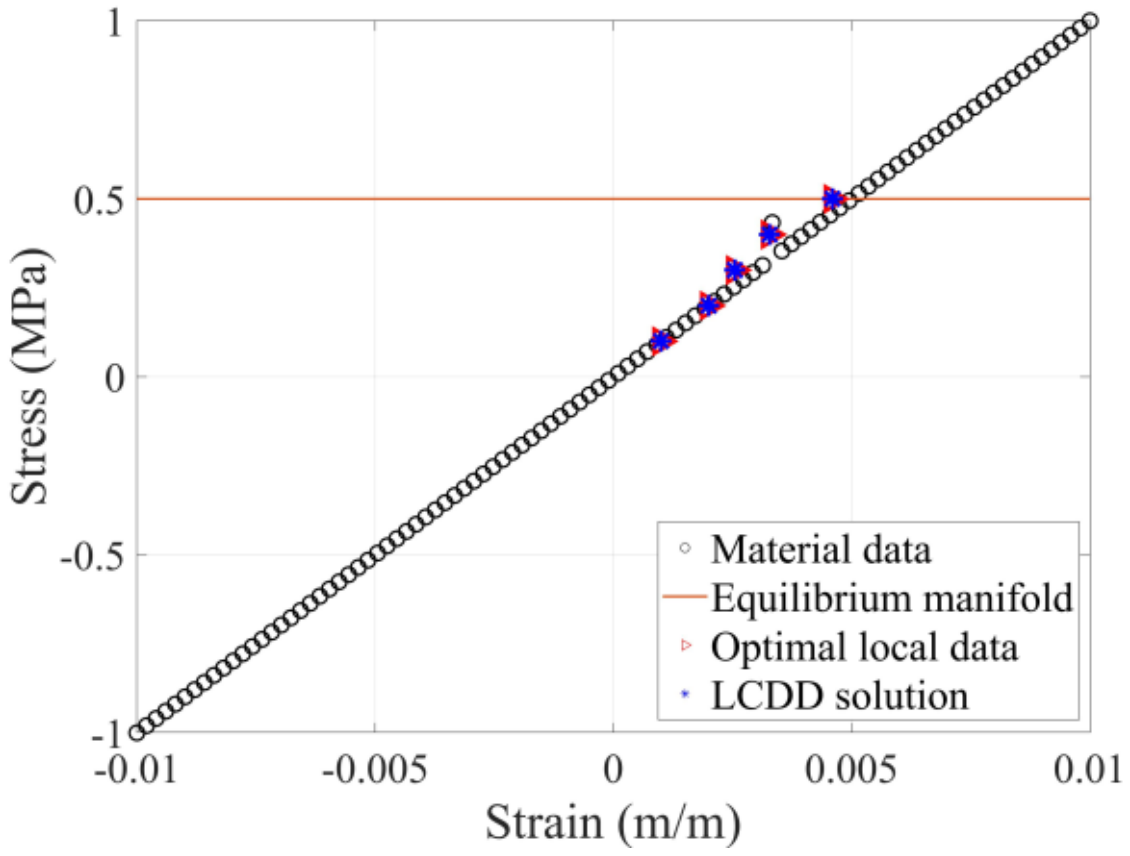

(c) LCDD: $k = 12$, $\bar{\mu} = 10^{-4}$

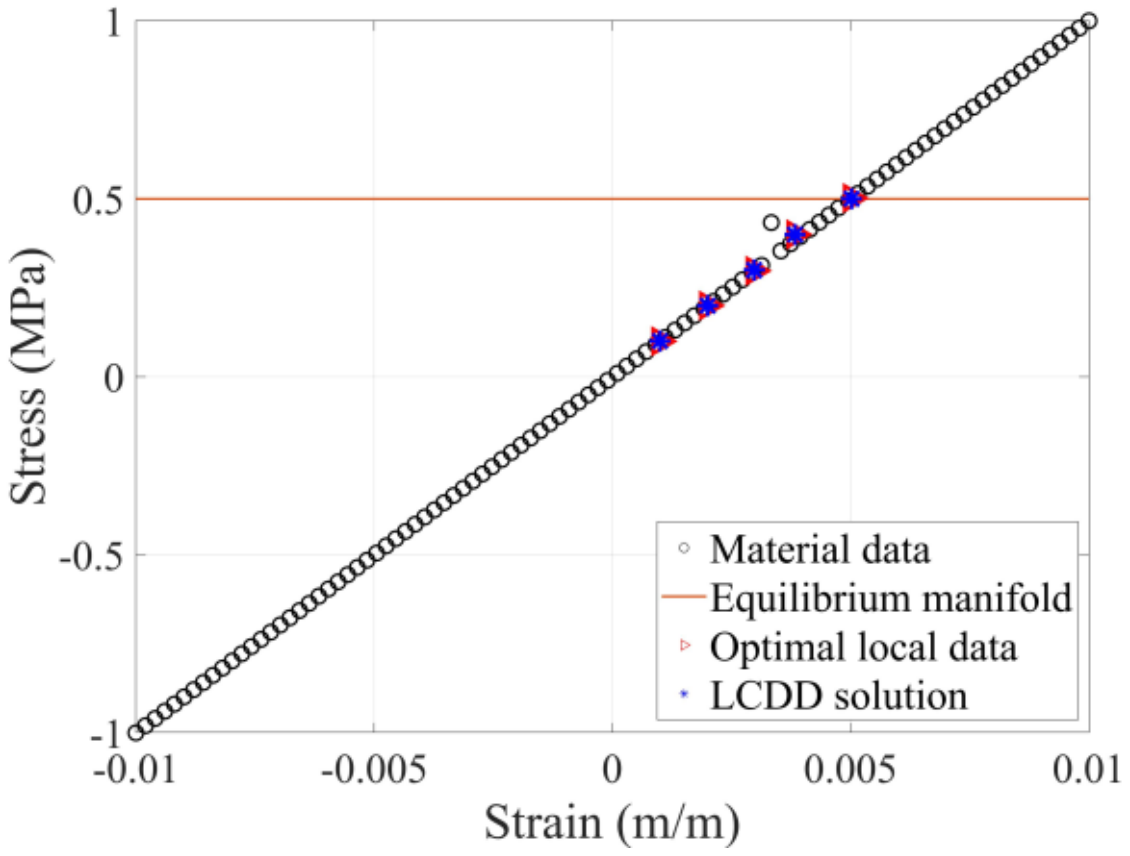

(d) LCDD: $k = 6$, $\bar{\mu} = 10^{-2}$

**Fig. 7.** Comparison of the data-driven solvers, DMDD and LCDD, for the one-bar truss structure using a material database with an outlier. The database contains $p = 100$ stress-strain data points. Different numbers of neighbor points $k$ and regularization coefficient values $\bar{\mu}$ are used in LCDD.

### *4.1.3. Data set characterizing a nonlinearly elastic material*

To reveal another pathology resulting from the discrete nature of data when using distance-minimizing approaches, consider a “nonlinear” database generated by a sigmoid function, as shown in **Fig. 8**. In the phase space, the plot of data points transits from a nearly linear stage of slope with $\sigma^i / \varepsilon^i = 100\text{Mpa}$ to a plateau at $\sigma = 0.51$ Mpa , which can be viewed as approximating a uniaxial perfect plasticity behavior.

**Fig. 8**a shows that although no noise is presented in this database, DMDD converges to a suboptimal solution before approaching the flat plateau, indicating the limitation of using discrete data space in the local solver (see (20)). This is because once the direction of data projection (linearly scaled with $M$ in this example) during data-driven iterations is nearly normal to the underlying material graph, the resulting displacement increment driven by the force residual (see (17a)) is too small to move the computational stresses and strains toward other data points in $\mathbb{E}$ that are closer to the physical manifold. As a result, the data-driven scheme converges at an undesirable solution. This issue is attributed to the non-continuous nature of discrete data, resulting in the susceptibility of DMDD to the selection of measure coefficient $M$ , the associated metric norm used to measure distance in the phase space, and the density and the underlying structure of data [54,60].

On the other hand, LCDD converges to a better solution (see **Fig. 8**b) at which the physical and material manifolds intersect. This is because when using the LCDD solver the inherited locally convex approximation represents a smooth constitutive (material) submanifold (i.e. the convex envelop) associated with the nonlinear material behavior. Since the locally convex reconstruction resembles the manifold learning technique introduced in [62] or local regression [74], LCDD is expected to reproduce a locally linear constitutive model corresponding to the sampled data points.

It should be emphasized that this linear reproducibility is very attractive in dealing with higher-dimensional phase space when data is relatively scarce, e.g., the elasticity problems in Section 5. As the reconstruction of local convexity confines the solution space for searching optimum local data in a bounded smooth domain, the proposed LCDD approach also avoids the non-convergence issue during the data-driven iterations, which usually appears in the regression based data-driven methods [47,55].

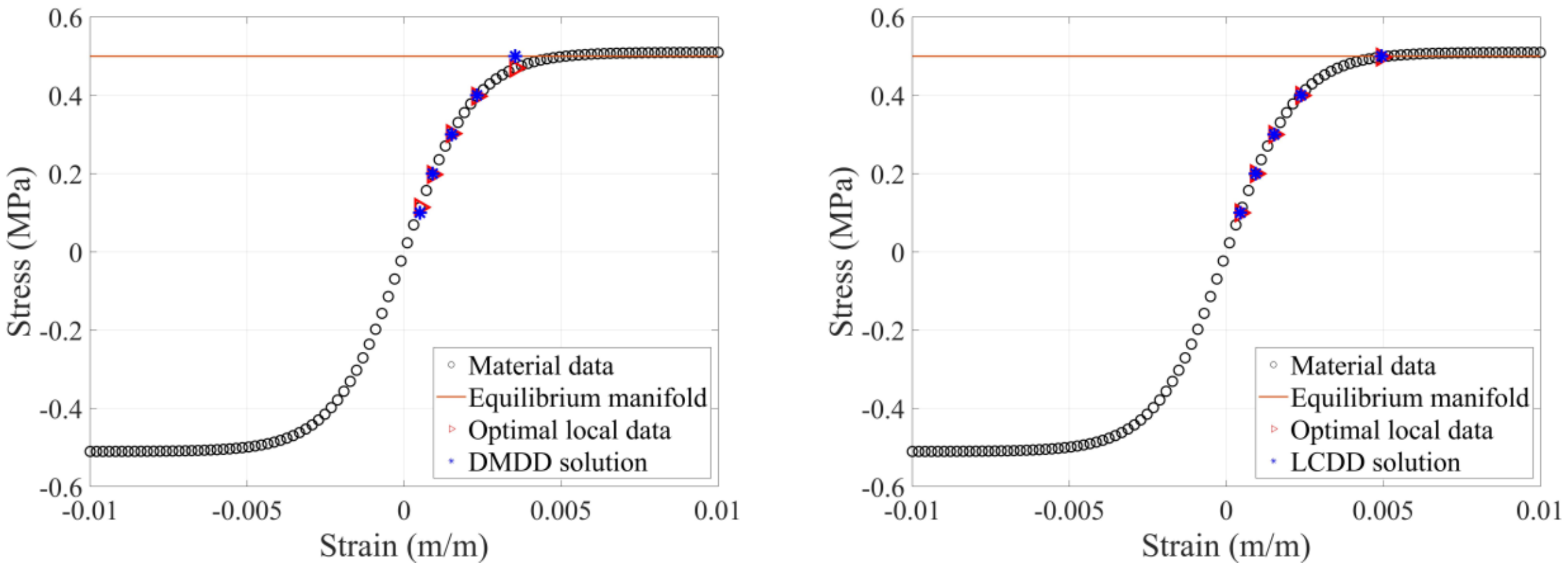

(a) (b)

**Fig. 8.** Comparison of the (a) DMDD and (b) LCDD solvers for the one-bar truss structure using a sigmoid material database. The database contains $p = 100$ stress-strain data points. The number of neighbor points used in LCDD is $k = 6$.

*4.2. Example II: Truss system*

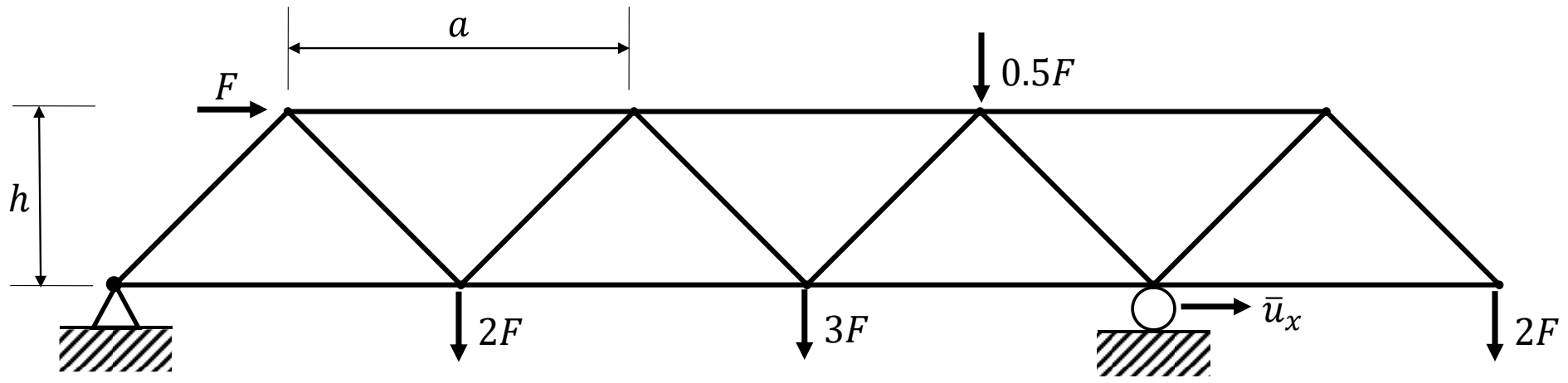

**Fig. 9.** A 15-bar truss structure with prescribed displacements and applied loads: $a = 4$ m, $h = 2$ m, $\bar{u}_x = 0.01$ m, and $F = 100$ kN.

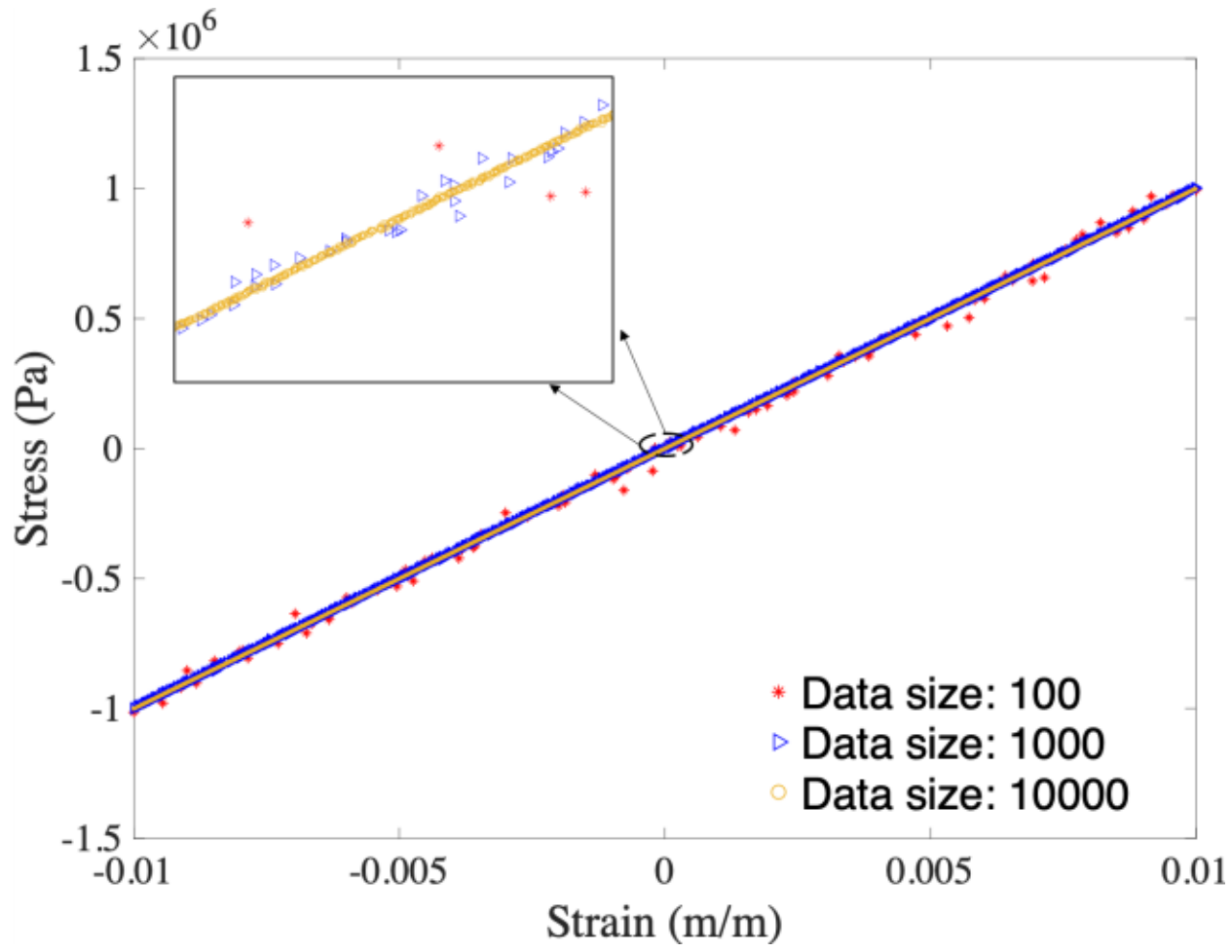

**Fig. 10.** Three noisy databases with different sizes, $p = 10^2, 10^3, 10^4$, where the random level of noise $\chi$ corresponding to each database is inverse to the data size, defined as $\chi = 2p^{-1}$.

To examine the convergence behavior with respect to the material data set size, consider a 15-bar truss structure (i.e., $m = 15$ for the local state vectors $\{\boldsymbol{s}_\alpha = [\varepsilon_\alpha \;\; \sigma_\alpha]^T\}_{\alpha=1}^m$) with unity cross-sectional area, as illustrated in **Fig. 9**. The solution obtained from different data-driven solvers are compared against the reference solution using the following normalized root-mean-square (%RMS) state errors

$$\varepsilon_{(\%\mathrm{RMS})} = \frac{1}{\varepsilon_{\max}^{\mathrm{ref}}} \left( \frac{1}{m} \sum_{\alpha=1}^{m} l_\alpha (\varepsilon_\alpha - \varepsilon_\alpha^{\mathrm{ref}})^2 \right)^{1/2}, \tag{33a}$$

$$\sigma_{(\%\mathrm{RMS})} = \frac{1}{\sigma_{\max}^{\mathrm{ref}}} \left( \frac{1}{m} \sum_{\alpha=1}^{m} l_\alpha (\sigma_\alpha - \sigma_\alpha^{\mathrm{ref}})^2 \right)^{1/2}, \tag{33b}$$

where $\{l_\alpha\}_{\alpha=1}^m$ are the length of the bars, $\{(\varepsilon_\alpha, \sigma_\alpha)\}_{\alpha=1}^m$ are the data-driven solutions for all bar members, $\{(\varepsilon_\alpha^{\mathrm{ref}}, \sigma_\alpha^{\mathrm{ref}})\}_{\alpha=1}^m$ are the strain and stress reference solutions corresponding synthetic material model, and $(\varepsilon_{\max}^{\mathrm{ref}}, \sigma_{\max}^{\mathrm{ref}})$ are the largest absolute values of strain and stress among all bar members.

In this numerical study, we consider three material data sets (see **Fig. 10**) with different sizes (i.e. $p = 10^2, 10^3, 10^4$), where the database with more data points is said to have higher density or larger data size, for the data-driven simulations. These noisy data sets are superimposed with the strain and stress perturbations given by the random Gaussian noise $\chi = 2p^{-1}$ (refer to (32)). In this case, the underlying structure of the data set uniformly converge to a linear curve with a slope of $M = 100$ MPa as the number of data points increases.

The convergence results of different data-driven solvers (DMDD and LCDD) measured by the normalized RMS state errors in (33) are shown in **Fig. 11** and **Fig. 12**. As suggested by the estimate given in [41,60], the data-driven solutions obtained by both methods converge towards the classical model-based solution with a rate close to 1 as the number of data points increases. However, a less satisfactory result is obtained by DMDD compared to LCDD. In **Fig. 11**, where LCDD yields nearly 1 order of accuracy higher than DMDD. This is due to the locally convex reconstruction that recovers the locally linear manifold. In LCDD, the inherent manifold learning ability contributes to the improved accuracy in addition to the enhancement of robustness against noisy data.

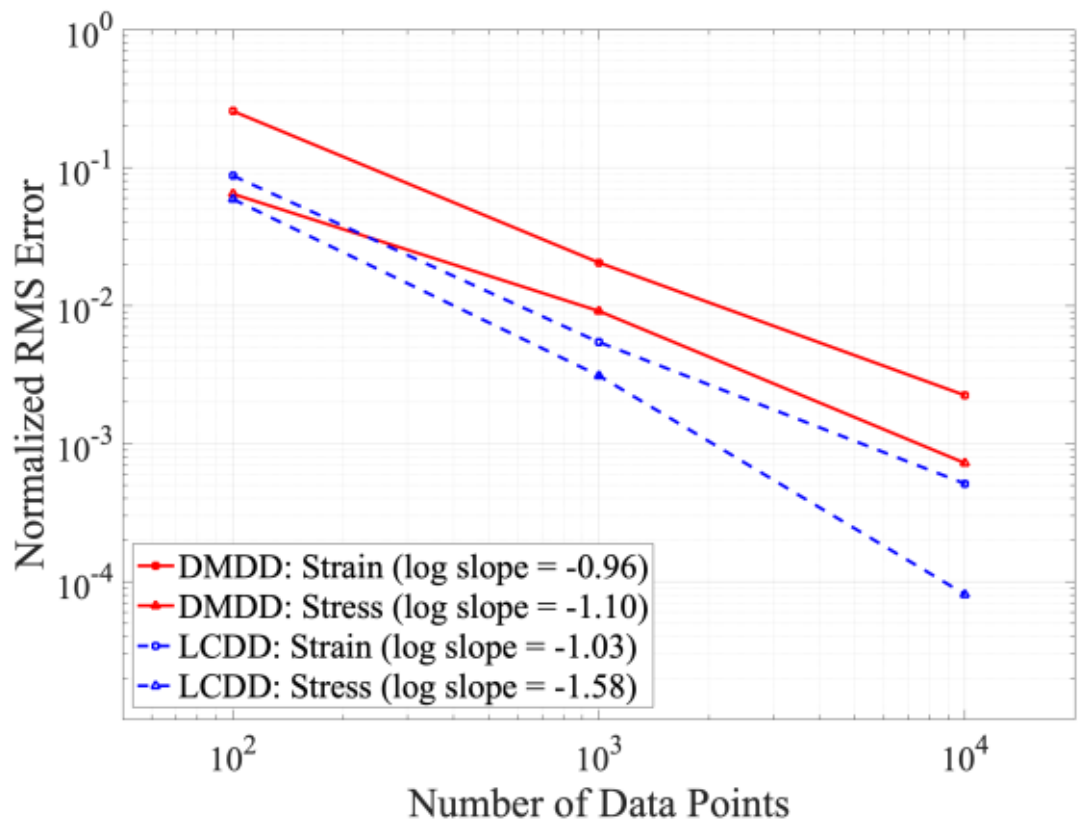

**Fig. 11.** Truss structure case. The convergence plot of the normalized RMS errors, $\varepsilon_{(\%\mathrm{RMS})}$ and $\sigma_{(\%\mathrm{RMS})}$, against increasing the size of database. The DMDD and LCDD solvers are compared using the three noisy data sets with different sizes, $p = 10^2, 10^3, 10^4$. The number of neighbor points used in LCDD is $k = 12$.

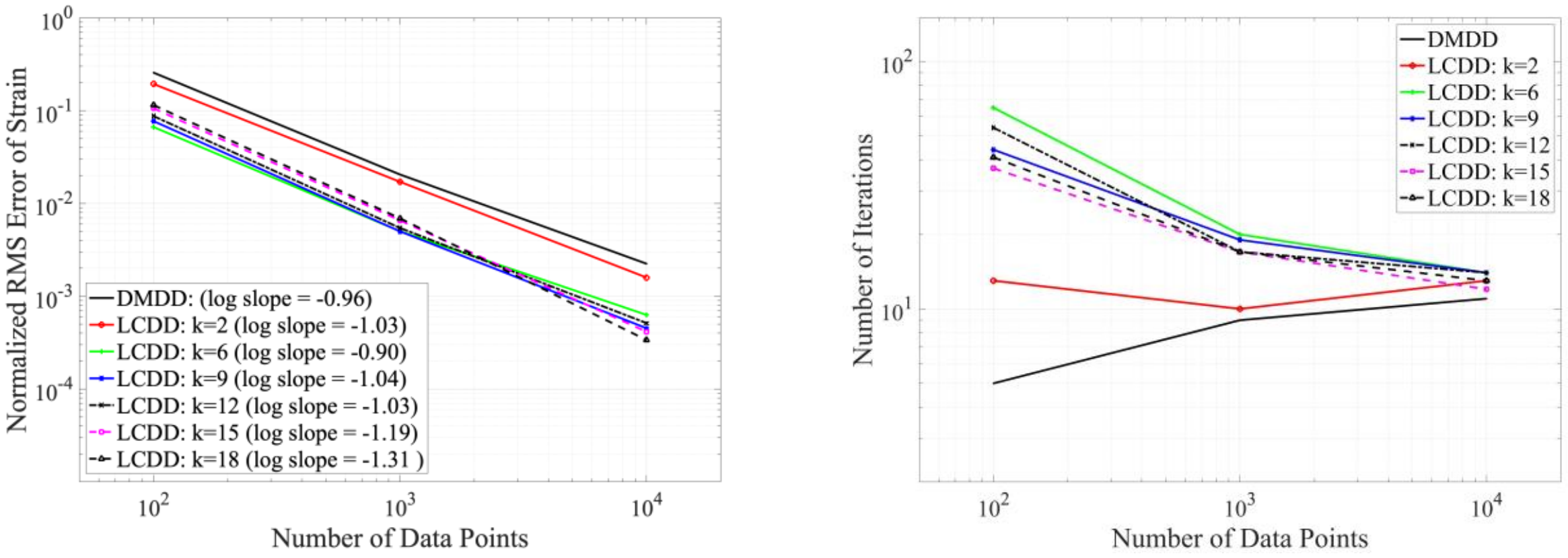

(a) (b)

**Fig. 12.** Truss structure case. (a) The convergence plot of the normalized RMS strain error $\varepsilon_{(\%\mathrm{RMS})}$ and (b) the number of iterations for convergence against increasing the size of database. The DMDD and LCDD solvers are compared using the noisy data sets with three different sizes, $p = 10^2, 10^3, 10^4$. Different number of neighbor points $k$ are used in LCDD.

It is also observed that the performance of the proposed LCDD solver appears to be insensitive to the numbers of convex hull neighbors (from $k = 6$ to $k = 18$) as demonstrated by its solution errors (**Fig. 12**a) and the number of iterations to attain convergence (**Fig. 12**b). Surprisingly, the results in **Fig. 12**b suggests that the LCDD solution converges faster as the data set size increases. This phenomenon is significantly distinct from other data-driven solvers, such as DMDD [41] and the max-ent data-driven solver [42], which require more iterations to achieve convergence when using larger data sets. We believe this is because the local manifold learning of LCDD better represents the underlying manifold. It is worth noting that when using $k = 2$, LCDD appears to lose accuracy and yields solutions approaching DMDD (shown in **Fig. 12**a), implying that LCDD would recover DMDD in the limit of using one neighbor.

A close comparison between the data-driven solutions of DMDD and LCDD using the data set with 100 material points (see **Fig. 10**) is given in **Fig. 13**. The reference solution (denoted by the diamond points) is obtained by utilizing the synthetic linear model (i.e. $\sigma = M\varepsilon$). As can be seen from **Fig. 13**a, the variations of the noisy data set substantially influence the DMDD performance such that several data-driven solutions (the asterisk points in a dashed box) converge poorly at some local minima that deviate from the linear graph, resulting in an overall unsatisfying performance of DMDD. In contrast, the LCDD solver overcomes such issues with noisy data as shown in **Fig. 13**b.

(a)

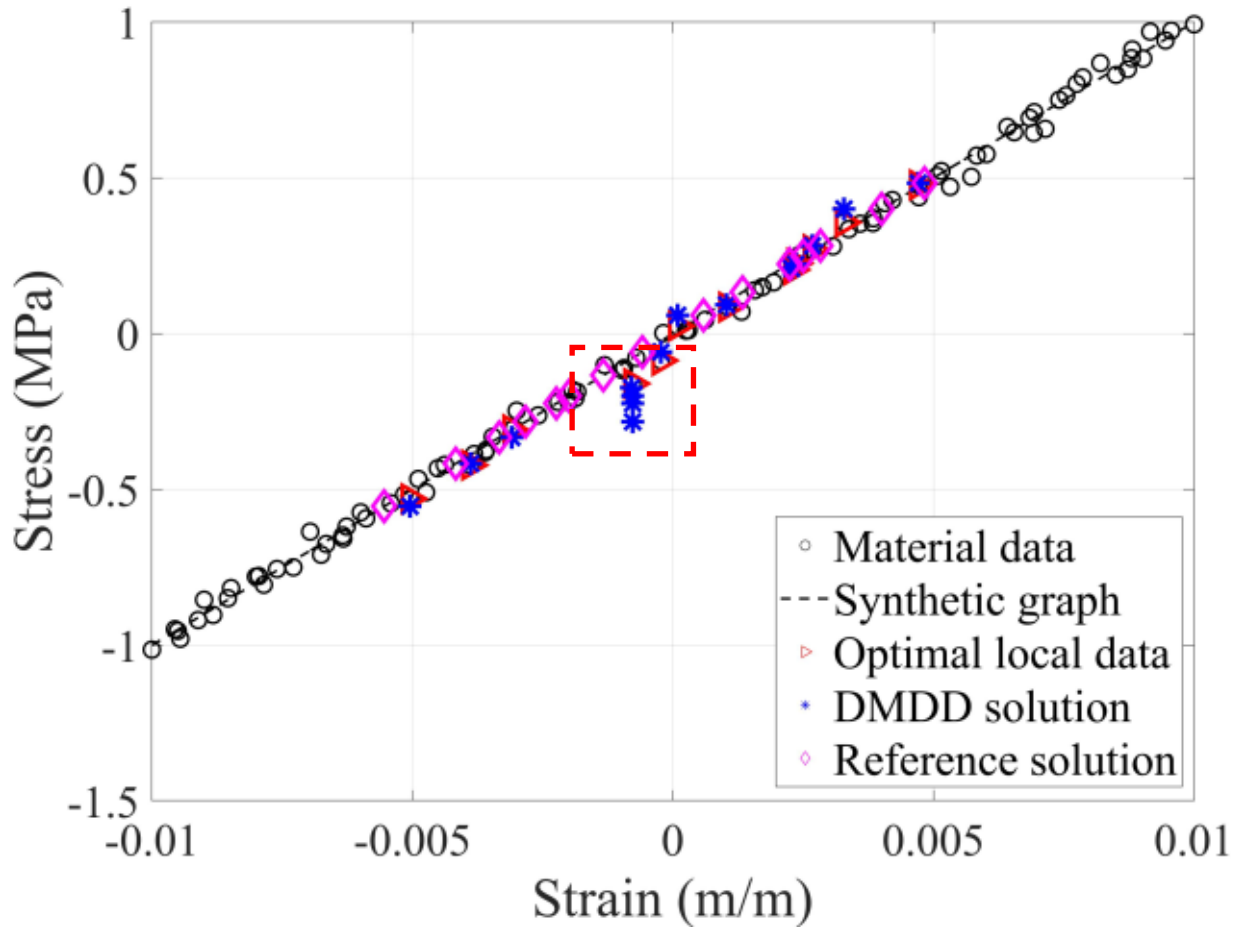

(b)

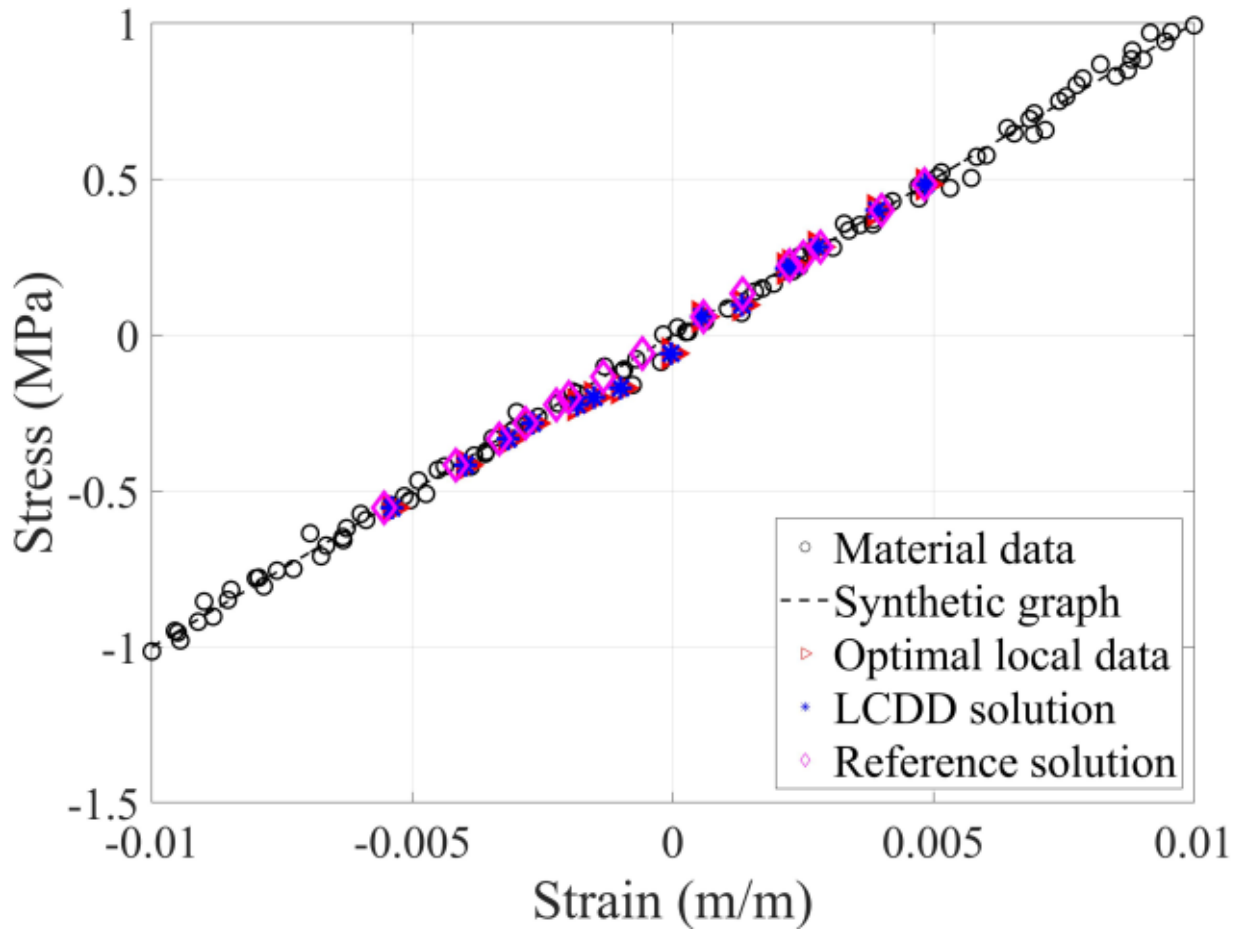

**Fig. 13.** Comparison of the data-driven solutions of (a) DMDD and (b) LCDD for the truss structure using the noisy material data set of $p = 100$ stress-strain data points. The number of neighbor points used in LCDD is $k = 12$.

## 5. Numerical Examples: Elasticity

In this section, we apply the proposed method to a two-dimensional elasticity problem. The accuracy and robustness of data-driven solvers are verified with high dimensionality of the associated phase space in the example problems.

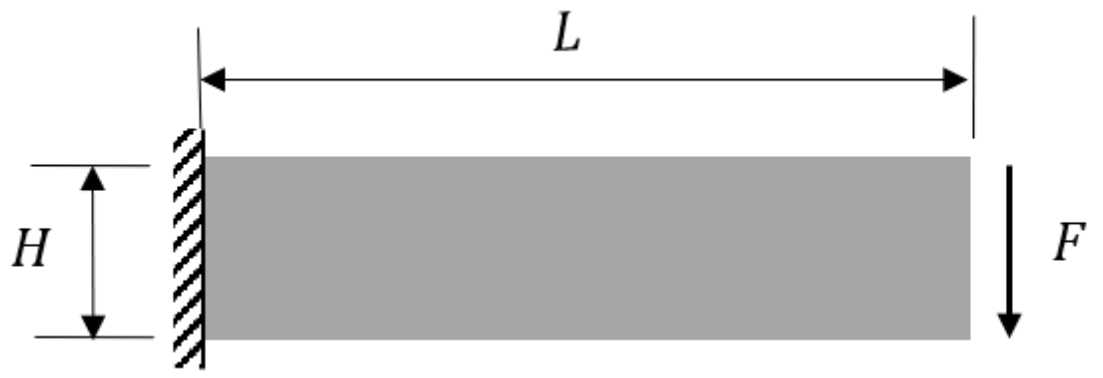

**Fig. 14.** A beam model subjected to a shear load: $L = 48\text{m}$, $H = 12\text{m}$ and $F = 1000\ \text{N}$.

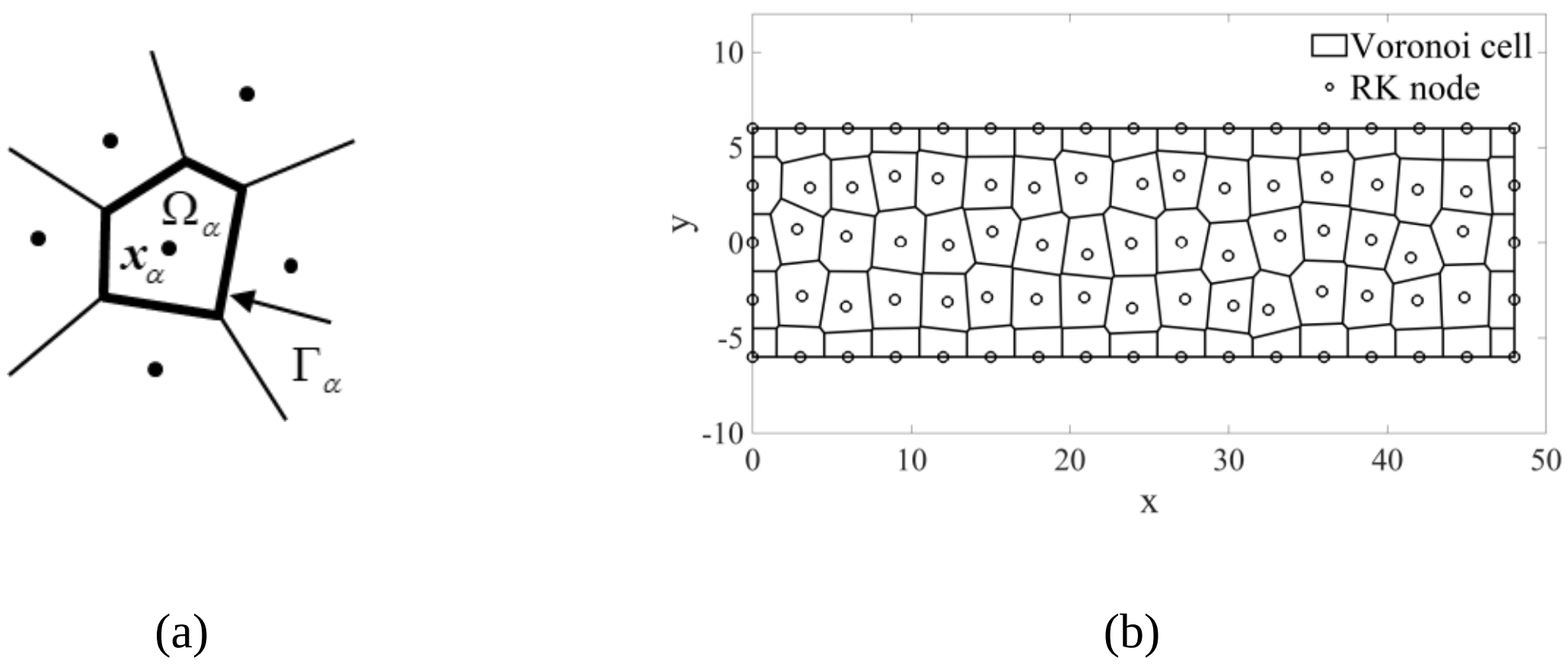

**Fig. 15.** Schematics of (a) Voronoi diagram and (b) the discretization of the beam model. The RK nodes are also the integration points under the SCNI framework.

A plane stress elastic beam subjected to a shear load as shown in **Fig. 14** is to be modeled. The reproducing kernel particle method (RKPM) with a stabilized conforming nodal integration (SCNI) approach are used for discretization, as shown in **Fig. 15**. More details about RKPM and SCNI can be referred to Appendix B. Under this RKPM-SCNI framework, the volumes of the Voronoi cells $\Omega_\alpha$, $\alpha = 1,...,N$, are used as the weights $V_\alpha$ in (16).

The synthetic elastic material model is given by the classical elasticity law with Young's modulus $E = 30$ MPa and Poisson's ratio $\nu = 0.3$. In this problem, the coefficient matrix used for norm (19) at each integration point is defined as

$$\boldsymbol{M} = \boldsymbol{M}_\alpha^\varepsilon = \boldsymbol{M}_\alpha^{\sigma-1} = \frac{E}{1-\nu^2}\begin{bmatrix} 1 & \nu & 0 \\ \nu & 1 & 0 \\ 0 & 0 & (1-\nu)/2 \end{bmatrix}. \tag{34}$$

To evaluate the performance of the data-driven solvers, the following normalized root-mean-square (%RMS) state error is defined for high-dimensional state,

$$\omega_{(\%\mathrm{RMS})} = \left( \frac{\sum_{\alpha=1}^{N} V_\alpha \| \boldsymbol{s}_\alpha - \boldsymbol{s}_\alpha^{\mathrm{ref}} \|_M^2}{\sum_{\alpha=1}^{N} V_\alpha \| \boldsymbol{s}_\alpha^{\mathrm{ref}} \|_M^2} \right)^{\frac{1}{2}}, \tag{35}$$

Where $\boldsymbol{s}_\alpha^{\mathrm{ref}} = [\boldsymbol{\varepsilon}_\alpha^{\mathrm{ref}\,T} \ \boldsymbol{\sigma}_\alpha^{\mathrm{ref}\,T}]^T$ denotes the nodal strain and stress reference solutions solved by using the synthetic material model, while $\boldsymbol{s}_\alpha = [\boldsymbol{\varepsilon}_\alpha^T \ \boldsymbol{\sigma}_\alpha^T]^T$ denotes the solutions solved by data-driven solvers using a given material data set.

Following the procedures of data generation in (32), the noiseless stress-strain data points using synthetic elastic material model in (34) are first generated, where the strains $\varepsilon_x$, $\varepsilon_y$ and $\gamma_{xy}$ are defined within the range $[-5\times10^{-4}, 5\times10^{-4}]$. The noisy data set is generated by superimposing the three components of noiseless strain and stress data, respectively, with the associated Gaussian noise term defined in (32), where $\chi = 0.4 \big/ \left(\sqrt[3]{p} l_{(j)}\right)$ and $l_{(j)}$ is the maximum value associated to $j$-th component of the noiseless data. Four material data sets in various size (i.e., $p = 10^3, 20^3, 40^3, 80^3$) are considered for the beam model.

The performance of the data-driven solvers using the noiseless data sets and the noisy data sets are given in **Fig. 16** and **Fig. 17**, respectively. Consistent to the convergence estimate in [41], the DMDD solutions converge linearly to the reference solution against the cubic root of the number of data points, regardless of using noiseless or noisy databases. LCDD using noiseless data sets (**Fig. 16**a) generates data-driven solution with the error as small as the convergence tolerance in the iterative data-driven computing (see Section 3.3). This implies that LCDD perfectly captures the underlying linear material graph even in such a high dimensional phase space. The convergence study with noisy data sets (**Fig. 17**a) shows that the LCDD solution using a sparse data set ( $p=10^3$ ) is able to achieve higher accuracy than the DMDD solution obtained using a very dense data set ( $p=80^3$ ), suggesting the superiority of LCDD over DMDD. Considering that it is difficult, in practice, to obtain a database with sufficiently dense data for high-dimensional spaces, the proposed LCDD approach is attractive.

As the intrinsic dimensionality of the employed linear elastic database is $d=2$, it is interesting to observe from **Fig. 16**a and **Fig. 17**a that the LCDD solutions obtained by using $k=3$ ( $d<k<2q$ , where $2q$ is the dimension of the material dataset $\mathbb{E}$ ) present an intermediate solution between the DMDD solution (i.e. $k=1$) and the other LCDD solutions with using more neighbor points $k \geq 2q=6$. The results indicate the importance of including enough neighbors in convex hull to fully preserve the manifold learning capacity in LCDD. In the case of noiseless data we observe that LCDD with $k=6$ is sufficient as its results are almost identical to the case with $k=9$ results(see **Fig. 16**).

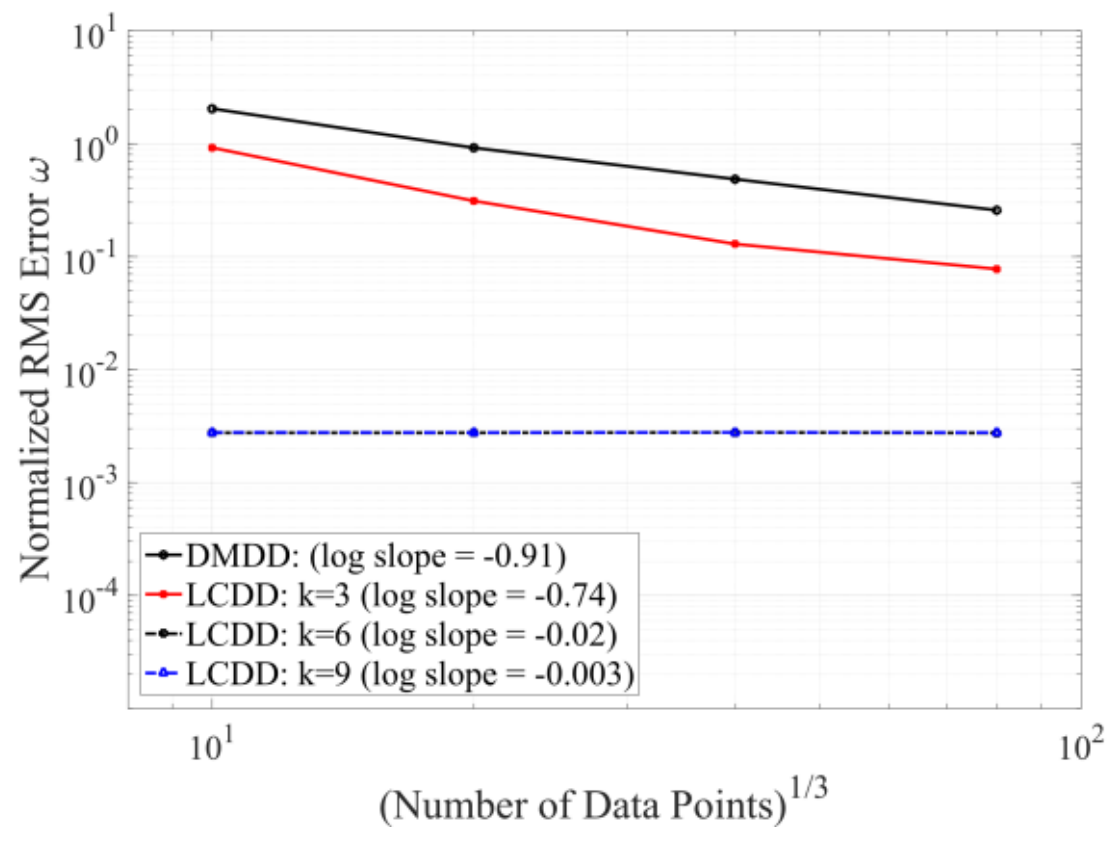

(a)

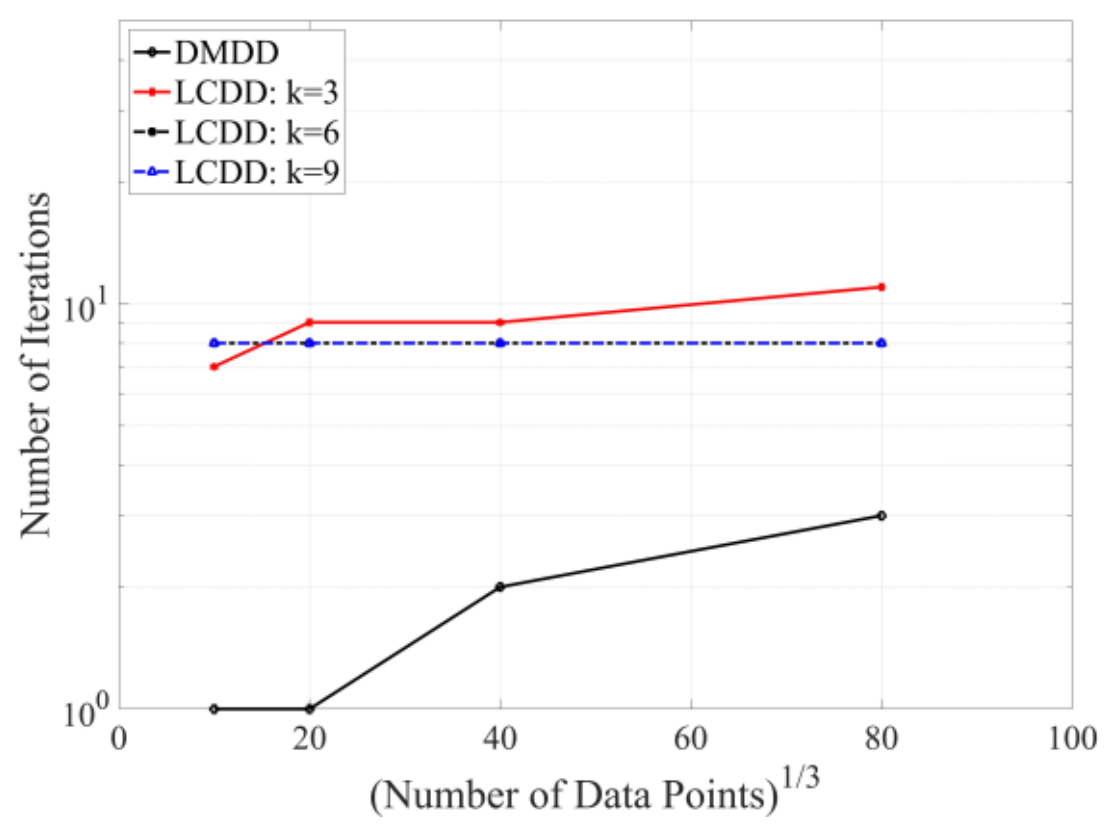

(b)

**Fig. 16.** Shear beam model with noiseless data sets. (a) The convergence plot of the normalized RMS state error $\omega_{(\%\mathrm{RMS})}$ and (b) the number of iterations for convergence against increasing the size of database. The DMDD and LCDD solvers are compared using the noiseless data sets with four different sizes, $p = 10^3, 20^3, 40^3, 80^3$. Different numbers of neighbor points $k$ are used in LCDD.

The associated number of convergence steps for the data-driven solvers are also presented in **Fig. 16**b and **Fig. 17**b. In contrast to the DMDD solver where the number of iterations increases with using a larger database, there is no evident increase for the LCDD solver. Moreover, the comparison of **Fig. 16**b and **Fig. 17**b shows that LCDD does not require more iterations between the local and global steps to converge when dealing with the noisy database rather than the noiseless database. It suggests that the convergence of LCDD is not sensitive to the size of database as well as the data sampling quality.

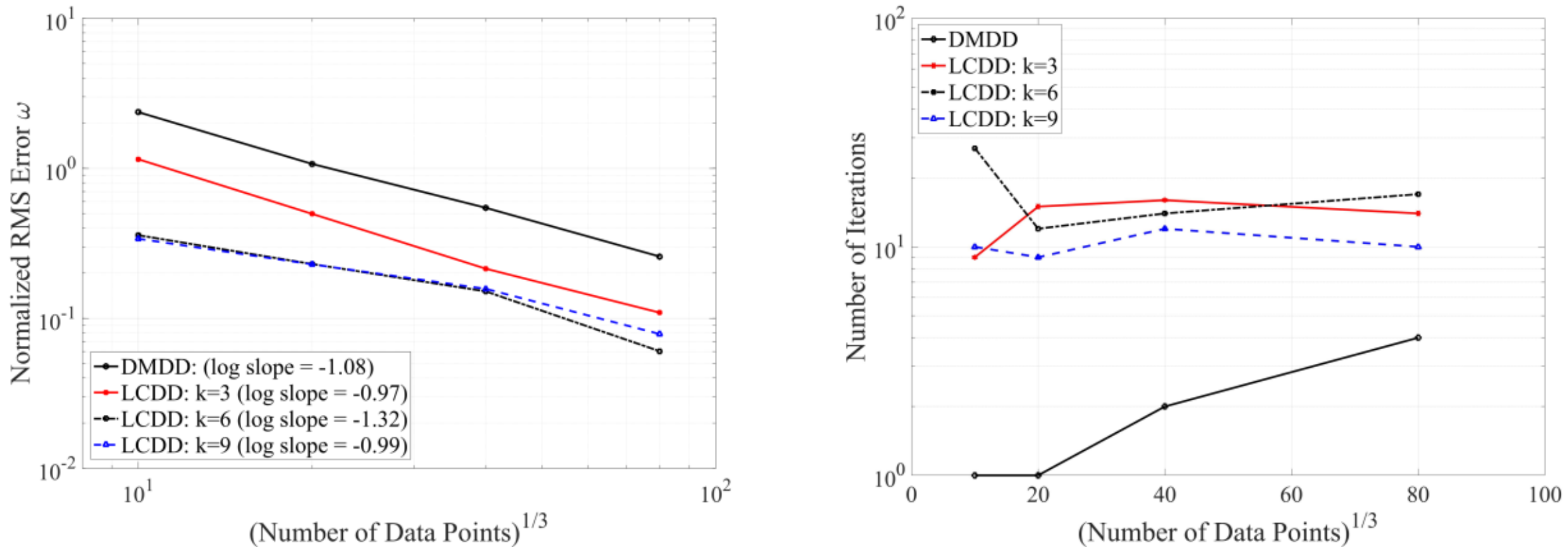

(a) (b)

**Fig. 17.** Shear beam model with noisy data sets. (a) The convergence plot of the normalized RMS state error $\omega_{(\%\mathrm{RMS})}$ and (b) the number of iterations for convergence against increasing the size of database. The DMDD and LCDD solvers are compared using the noisy data sets with four different sizes, $p = 10^3, 20^3, 40^3, 80^3$. Different numbers of neighbor points $k$ are used in LCDD.

The beam deformations simulated by the data-driven solvers are also compared in **Fig. 18** and **Fig. 19**. It is observed that DMDD performs poorly (**Fig. 18**a and **Fig. 19**a) due to the susceptibility to noisy data and local minimum issues that are more pronounced in elasticity problems with a high-dimensional phase space. On the other hand, LCDD exactly reproduces the reference solutions when using noiseless database (**Fig. 18**b), and results in marginal deviations from the reference when using the noisy database (**Fig. 19**b). Moreover, **Fig. 20** shows the stress solutions, $\sigma_{xx}$ and $\sigma_{xy}$, obtained by LCDD compared to the constitutive model-based reference solutions. It shows that LCDD can yield accurate stress solutions across the problem domain with using a noisy database. This demonstrates that the LCDD approach remains robust with noisy data in solving elasticity problems.

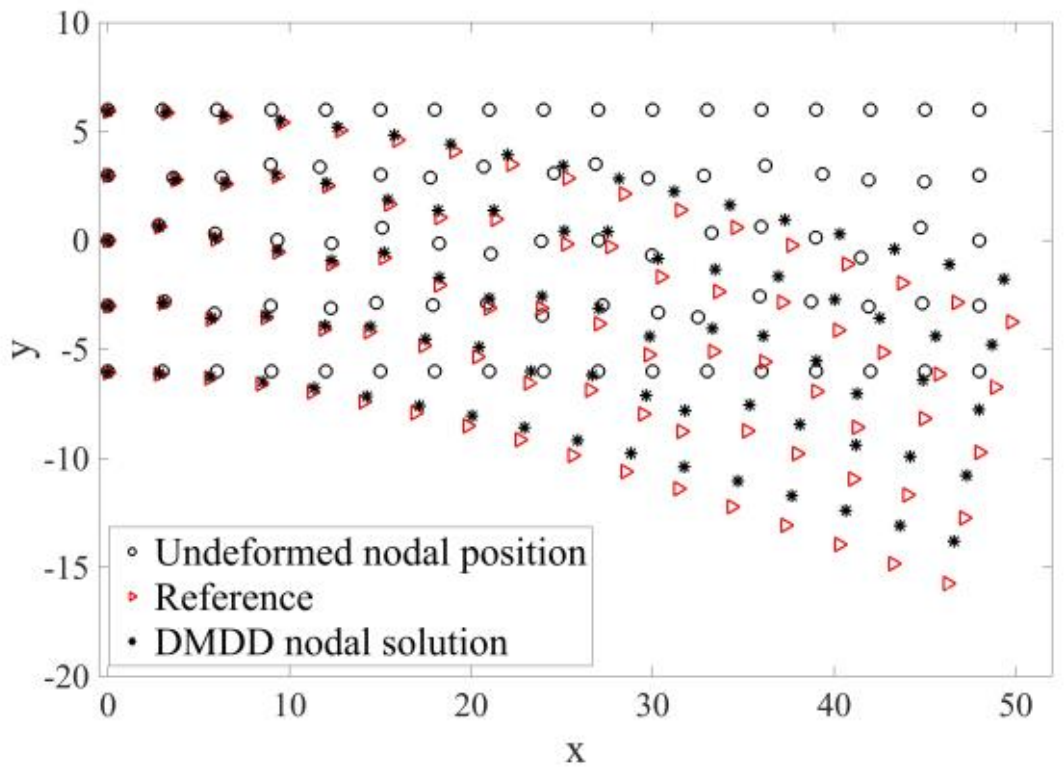

(a)

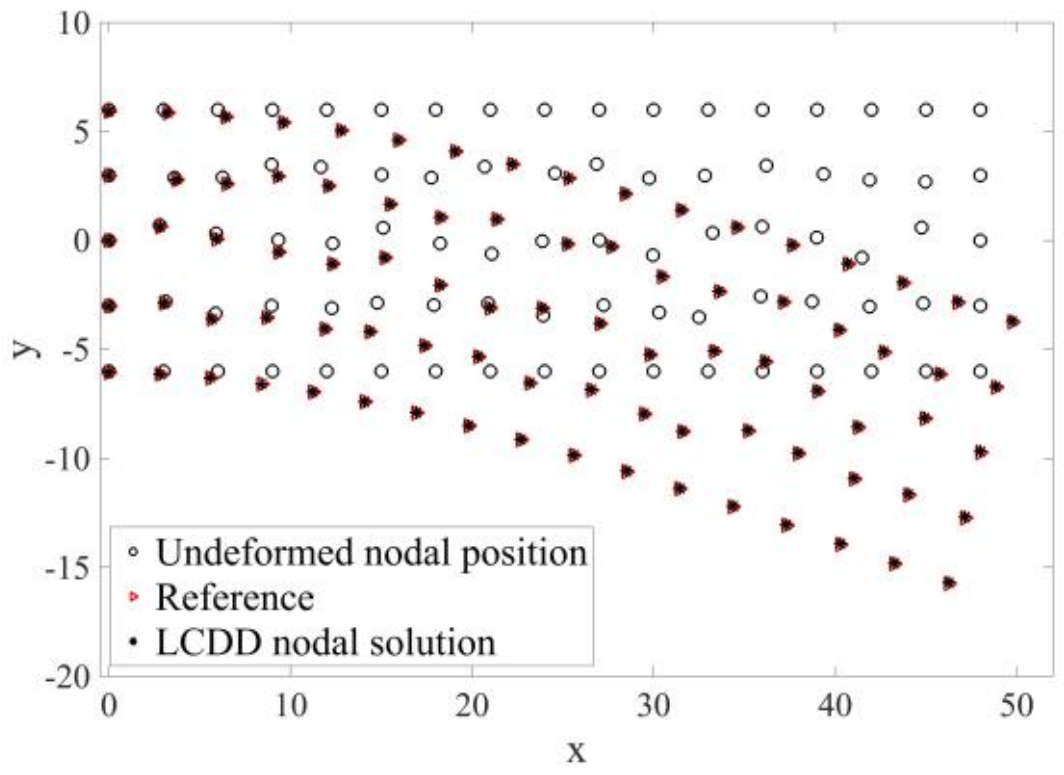

(b)

**Fig. 18.** Comparison of the data-driven displacement solutions for the shear beam model by using (a) DMDD with a noiseless data set of $p = 80^3$ stress-strain data points and (b) LCDD with a noiseless data set of $p = 10^3$ stress-strain data points. The number of neighbor points used in LCDD is $k = 6$.

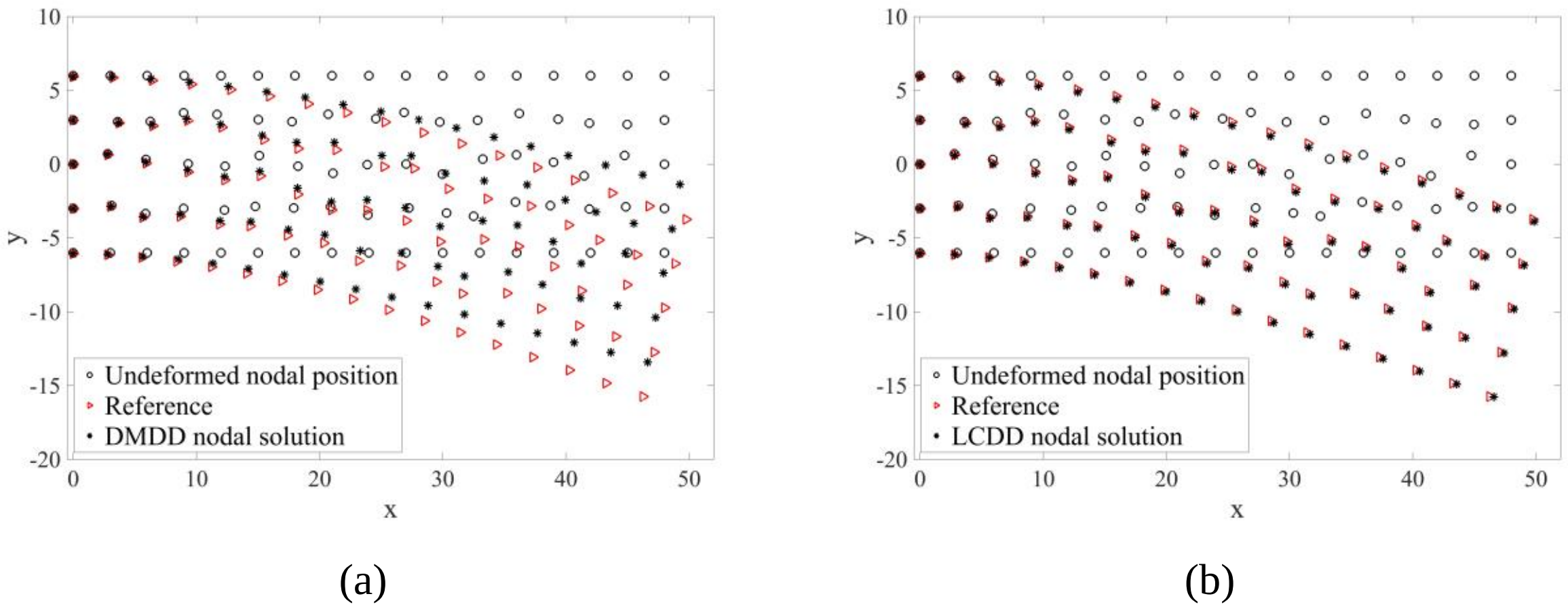

**Fig. 19.** Comparison of the data-driven displacement solutions for the shear beam model by using (a) DMDD with a noisy data set of $p = 80^3$ stress-strain data points and (b) LCDD with a noisy data set of $p = 10^3$ stress-strain data points. The number of neighbor points used in LCDD is $k = 6$

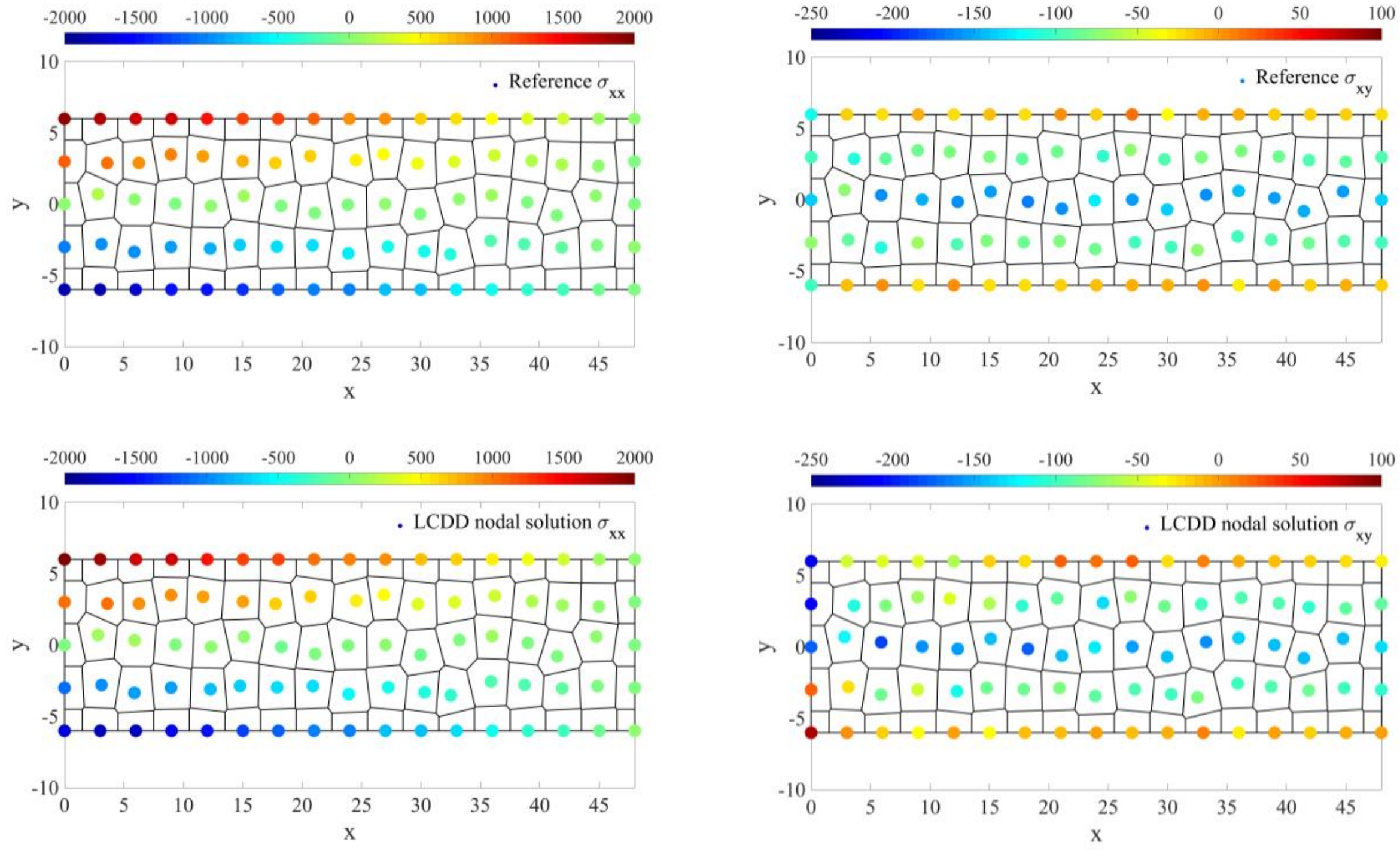

**Fig. 20.** Comparison of the stress solutions, $\sigma_{xx}$ and $\sigma_{xy}$, of the reference constitutive model-based computing (upper) and the proposed LCDD computing (bottom) for the shear beam model. A noisy data set of $p = 10^3$ stress-strain data points and $k = 6$ neighbor points are used in LCDD.

## 6. Discussion & Conclusion

We have formulated a local convexity data-driven (LCDD) solver as a new data-driven computing paradigm integrated with manifold learning techniques that generalizes the distance-minimizing data-driven (DMDD) computing [41] with enhanced accuracy and robustness of data-driven computing against noise/outliers. The proposed method adaptively selects $k$ nearest neighbor ($k$-NN) material data points for physical local states to be updated and searches for optimal data solutions from a bounded solution space defined by the convex hull of the selected $k$-NN points. This local data searching procedure has been formulated under a *non-negative least squares* (NNLS) solver and can be solved efficiently. By means of the clustering analysis based on $k$-NN and preserving the convexity of the constructed local material manifold, LCDD yields enhanced robustness and convergence stability.

From the *pure data-driven approach* point of view (refer to [54]), LCDD is inspired by measuring the distance to a local convex set $\mathcal{E}$ , instead of from a single discrete data, aiming to enhance the robustness against noise and prevent undesirable local minima. LCDD can be reduced to the standard DMDD approach when using only one neighbor data point, i.e. $k=1$. Thus, LCDD retains the simplicity and is computationally efficiency compared to other robustness enhanced data-driven methods with introducing statistics information [42,54]. From the *fitted data-driven (or linearization) approach* point of view, on the other hand, LCDD relies on the approximation of locally linear material graph by the manifold learning methodologies [21,63] to capture the global structure via local data information. However, the proposed LCDD scheme distinguishes itself from the other manifold learning based data-driven approaches [47, 55] with the following two aspects: first, the iteration process of LCDD does not need to explicitly construct constitutive manifold and use tangent information; second, LCDD introduces the convexity condition on the reconstructed material graph, thus avoiding convergence issues that occur when using standard regression approaches. In addition, we believe preserving convexity is also of physical importance. For example, it is expected to better preserve the positivity of strain energy via LCDD rather than other manifold learning techniques because the proposed approach learns the underlying constitutive manifold based on a convex combination of a cluster of neighboring data points, although a rigorous analysis of this question is beyond the scope of this paper.

It has also been shown that the embedded NNLS solver seeks the projection point of a given computational state onto a nearby material graph implicitly constructed based on the *k*-NN points, which ensures local linear reproducibility in the approximation. Hence, in addition to the improved robustness and accuracy in dealing with noisy database, LCDD exactly represent the underlying linear stress-strain relationship. In the proposed global-local data-driven algorithm, smooth solution spaces are employed for the global physical solution and the local project on the convex set. This is achieved by introducing a RK shape functions in the approximation of the global physics laws and the regularized LCDD learning solver in the optimal data set search. With the SCNI domain integration employed in the global Galerkin equations, it significantly reduces the needed stress-strain data search in the local LCDD learning solver, leading to an effective data-driven computing. The proposed LCDD dada-driven method has been applied to truss problems with linear and nonlinear stress-strain relationship, and continuum elasticity problems and demonstrated its effectiveness in robustness, convergence, accuracy in high-dimensional phase spaces.

This paper is intended to introduce manifold learning techniques, or dimensionality reduction [21], to data-driven computing. Our numerical studies show that it is effective in applying manifold learning for problems with high-dimensional data, because in high-dimensional spaces the data can be extremely sparse and the acquisition of sufficient data is not practical. This demands effective dimensionality reduction to identify and extract the essential information from the database, and elasticity example in Section 5 demonstrates the suitability of LCDD with inherent manifold learning for such problems.

Although manifold learning has been shown to enhance the convergence performance during data-driven iteration, the computational cost of each local step remains linearly scaled with the size of material database. Thus, when the datasets become large with high-dimensional information, such as time-dependent states [43], inelastic quantities [35,50,51,82], and so on, more advanced machine learning models for manifold learning like autoencoder can be employed. This is the direction of our further research.

## Acknowledgements

The support of this work by the National Science Foundation under Award Number CCF-1564302 to University of California, San Diego, is greatly appreciated.

## Appendix A. Non-negative least squares solver

Let us recall a standard non-negative least squares (NNLS) problem: *given a matrix* $\boldsymbol{A} \in \mathbb{R}^{n \times p}$ *(usually* $p > n$*) and a observed vector* $\boldsymbol{z} \in \mathbb{R}^{n}$*, find a nonnegative vector* $\boldsymbol{y}^{*} \in \mathbb{R}_{+}^{p}$ *to minimize the following function,*

$$\begin{aligned} &\boldsymbol{y}^{*} = \arg\min_{\boldsymbol{x} \in \mathbb{R}^{p}} \| \boldsymbol{A}\boldsymbol{y} - \boldsymbol{z} \|, \\ &\text{subject to: } y_i \geq 0,\ i = 1, \ldots, p, \end{aligned} \tag{A.1}$$

*where* $\| \cdot \|$ *stands for the standard Euclidean norm.*

A variety of methods have been applied to tackle the NNLS problem since 1980s. Those algorithms in general can be roughly categorized into active-set methods and iterative approaches [75]. Lawson and Hanson [56] seems to propose the first standard algorithm to solve NNLS problem (A.1). Their method is essentially an active set method [76], which is based on the observation that only a small subset of the non-negative constraints are usually active at the solution. It shows in [56] that the iteration in the active set method converges and terminates without any cutoff in iterations. The standard algorithm for active-set method is reviewed in Algorithm 1.

**Algorithm 1** Non-negative least squares solver: $\boldsymbol{y}^* \leftarrow NNLS(\boldsymbol{A}, \boldsymbol{z}, TOL)$

**Input**: $\boldsymbol{A} \in \mathbb{R}^{n \times p}$, $\boldsymbol{z} \in \mathbb{R}^n$, TOL

**Output**: $\boldsymbol{y}^* \succ= \mathbf{0}$ such that $\boldsymbol{y}^* = \arg\min \| \boldsymbol{A}\boldsymbol{y} - \boldsymbol{z} \|^2$

**Initialization**: $\mathcal{Z} \leftarrow \emptyset$, $\mathcal{Y} \leftarrow \{1, 2, \ldots, p\}$, $\boldsymbol{y} = \boldsymbol{0}$, $\boldsymbol{r} \leftarrow \boldsymbol{z}$

WHILE $\| \boldsymbol{r} \| / \| \boldsymbol{z} \| > TOL$ and $\mathcal{Y} \neq \emptyset$, DO

  $\boldsymbol{q} \leftarrow \boldsymbol{A}^T(\boldsymbol{z} - \boldsymbol{A}\boldsymbol{y})$, $j = \arg\max_{i=1,\ldots,p}(q_i)$

  Include the index $j$ in $\mathcal{Z}$ and remove it from $\mathcal{Y}$

  $\boldsymbol{s}_{\mathcal{Z}} \leftarrow (\boldsymbol{A}_{\mathcal{Z}}^T \boldsymbol{A}_{\mathcal{Z}})^{-1} \boldsymbol{A}_{\mathcal{Z}}^T \boldsymbol{z}$, $\boldsymbol{s}_{\mathcal{Y}} \leftarrow \boldsymbol{0}$, where $\boldsymbol{A}_{\mathcal{Z}} \in \mathbb{R}^{n \times |\mathcal{Z}|}$

  WHILE $\min_{i \in \mathcal{Z}} \boldsymbol{s}_{\mathcal{Z}} \leq 0$, DO

    $\alpha = -\min_{i \in \mathcal{Z}} y_i / (y_i - s_i)$

    $\boldsymbol{y} \leftarrow \boldsymbol{y} + \alpha(\boldsymbol{s} - \boldsymbol{y})$

    Update $\mathcal{Y}$ with zero value indices of $\boldsymbol{y}$ and $\mathcal{Z}$ with the positive indices of $\boldsymbol{y}$

    $\boldsymbol{s}_{\mathcal{Z}} \leftarrow (\boldsymbol{A}_{\mathcal{Z}}^T \boldsymbol{A}_{\mathcal{Z}})^{-1} \boldsymbol{A}_{\mathcal{Z}}^T \boldsymbol{z}$, $\boldsymbol{s}_{\mathcal{Y}} \leftarrow \boldsymbol{0}$

  $\boldsymbol{y} \leftarrow \boldsymbol{s} = \boldsymbol{s}_{\mathcal{Z}} \cup \boldsymbol{s}_{\mathcal{Y}}$

  $\boldsymbol{r} \leftarrow \boldsymbol{z} - \boldsymbol{A}\boldsymbol{y}$

Return $\boldsymbol{y}^* \leftarrow \boldsymbol{y}$

## Appendix B. Meshfree approximation

In the section we review the reproducing kernel paritile method (RKPM) [57,58] and the stabilized conforming nodal integration (SCNI) approach [59] that are used for solving the weak form in (13).

### *B.1. Reproducing kernel approximation*

The reproducing kernel (RK) shape functions $\{\Psi_I\}_{I=1}^{N}$ used for approximating displacement and Lagrange multipliers in (14) are expressed as

$$\Psi_I(\boldsymbol{x}) = C(\boldsymbol{x};\boldsymbol{x}-\boldsymbol{x}_I)\Phi_a(\boldsymbol{x}-\boldsymbol{x}_I) \quad \text{(B.1)}$$

The kernel function $\Phi_a$ defines a local support for the shape functions by a support size "$a$" as well as the smoothness of the approximation. A widely used kernel function is the cubic-B splines that provides $C^2$ continuity, expressed as

$$\Phi_a(\boldsymbol{x}-\boldsymbol{x}_I) = \Phi_a(z) = \begin{cases} 2/3 - 4z^2 + 4z^3 & \text{for } 0 \le z < 1/2 \\ 4/3 - 4z + 4z^2 - 4/3\, z^3 & \text{for } 1/2 \le z < 1 \\ 0 & \text{for } z \ge 1 \end{cases} \quad \text{(B.2)}$$

where $z = \| \boldsymbol{x} - \boldsymbol{x}_I \| / a$. The term $C(\boldsymbol{x};\boldsymbol{x}-\boldsymbol{x}_I)$ is a correction function constructed using a set of basis functions,

$$C(\boldsymbol{x};\boldsymbol{x}-\boldsymbol{x}_I) = \sum_{i+j+k=0}^{n} (x_1 - x_{1I})^i (x_2 - x_{2I})^j (x_3 - x_{3I})^k b_{ijk}(\boldsymbol{x}) = \boldsymbol{H}^T(\boldsymbol{x}-\boldsymbol{x}_I)\boldsymbol{b}(\boldsymbol{x}) \quad \text{(B.3)}$$

in which $\boldsymbol{H}(\boldsymbol{x}-\boldsymbol{x}_I)$ is a vector consisting of all the monomial basis functions upto *n*-th order, and $\boldsymbol{b}$ is an unknown parameter vector determined by enforcing the *n*-th order reproducing conditions as follows,

$$\sum_{I=1}^{N} \Psi_I(\boldsymbol{x}) x_{1I}^i x_{2I}^j x_{3I}^k = x_1^i x_2^j x_3^k, \quad |i+j+k| = 0,1,\ldots,n \quad \text{(B.4)}$$

Introducing Eqs. (B.1) and (B.3) into (B.4), the coefficient vector can be obtained by

$$\boldsymbol{b}(\boldsymbol{x}) = \boldsymbol{M}^{-1}(\boldsymbol{x})\boldsymbol{H}(\boldsymbol{0}) \quad \text{(B.5)}$$

where the moment matrix is

$$\boldsymbol{M}(\boldsymbol{x}) = \sum_{I=1}^{N} \boldsymbol{H}(\boldsymbol{x} - \boldsymbol{x}_I)\boldsymbol{H}^T(\boldsymbol{x} - \boldsymbol{x}_I)\Phi_a(\boldsymbol{x} - \boldsymbol{x}_I) \tag{B.6}$$

Finially, the RK shape functions is expressed as,

$$\Psi_I(\boldsymbol{x}) = \boldsymbol{H}^T(\boldsymbol{0})\boldsymbol{M}^{-1}(\boldsymbol{x})\boldsymbol{H}(\boldsymbol{x} - \boldsymbol{x}_I)\Phi_a(\boldsymbol{x} - \boldsymbol{x}_I) \tag{B.7}$$

It should be noted that the above RK shape functions do not possess the Kronecker delta property. Thus, certain techniques are needed to impose kinematically admissible approximation on the essential boundary, such as Lagrange multipliers method [77], Nitsche's method [78,79], boudary singular kernel method [80], and transformation methods [58,80]. More discussions on the mathematical properties of the reproducing kernel approaximation can be found in the review paper [81].

*B.2. Nodal integration scheme*

The SCNI approach is employed for the domain integration of the weak form (13) to achieve computational efficiency and accuracy when using meshfree shape functions with nodal integration quadrature schemes.

The key idea behind SCNI is to satisfy the linear patch test (thus, ensure the linear consistency) by leveraging a condition, i.e. the divergence constraint on the test function space and numerical integration [59], expressed as:

$$\hat{\int_{\Omega}} \nabla \Psi_I \mathrm{d}\Omega = \hat{\int_{\partial\Omega}} \Psi_I \boldsymbol{n} \mathrm{d}\Gamma, \tag{B.8}$$

where '^' over the integral symbol denotes numerical integration. In SCNI, an effective way to achieve Eq. (B.8) is based on nodal integration with gradients smoothed over conforming representative nodal domains, as shown in **Fig. 21**, converted to boundary integration using the divergence theorem

$$\tilde{\nabla} \Psi_I(\boldsymbol{x}_L) = \frac{1}{V_L}\int_{\Omega_L} \nabla \Psi_I \mathrm{d}\Omega = \frac{1}{V_L}\int_{\partial\Omega_L} \Psi_I \boldsymbol{n} \mathrm{d}\Gamma, \tag{B.9}$$

where $V_L = \int_{\Omega_L} d\Omega$ is the volume of a conforming smoothing domain associated to the node $\boldsymbol{x}_L$, and $\tilde{\nabla}$ denotes the smoothed gradient operator. In this method, smoothed gradients are employed for both test and trial functions, as the approximation in (B.9) enjoys first order completeness and leads to a quadratic rate of convergence for solving linear solid problems by meshfree Galerkin methods. As shown in **Fig. 21**, the continuum domain $\Omega$ is partitioned into $N$ conforming cells by Voronoi diagram, and both the nodal displacement vectors and the state variables (e.g., stress, strain) are defined at the set of nodes at $\{\boldsymbol{x}_L\}_{L=1}^{N}$.

Therefore, if we consider two-dimensional elasticity problem under the SCNI framework, the smoothed strain-displacement matrix $\tilde{\boldsymbol{B}}_I(\boldsymbol{x}_L)$ used in (16) is expressed as:

$$\tilde{\boldsymbol{B}}_I(\boldsymbol{x}_L) = \begin{bmatrix} \tilde{b}_{I1}(\boldsymbol{x}_L) & 0 \\ 0 & \tilde{b}_{I2}(\boldsymbol{x}_L) \\ \tilde{b}_{I2}(\boldsymbol{x}_L) & \tilde{b}_{I1}(\boldsymbol{x}_L) \end{bmatrix}, \tag{B.10}$$

$$\tilde{b}_{Ii}(\boldsymbol{x}_L) = \frac{1}{V_L}\int_{\partial\Omega_L} \Psi_I(\boldsymbol{x}) n_i(\boldsymbol{x}) \mathrm{d}\Gamma. \tag{B.11}$$

Since the employment of the smoothed gradient operator in (B.9) and (B.11) satisfies the divergence constraint regardless of the numerical boundary integration, a two trapezoidal rule for each segment of $\partial\Omega_L$ is used in this study.

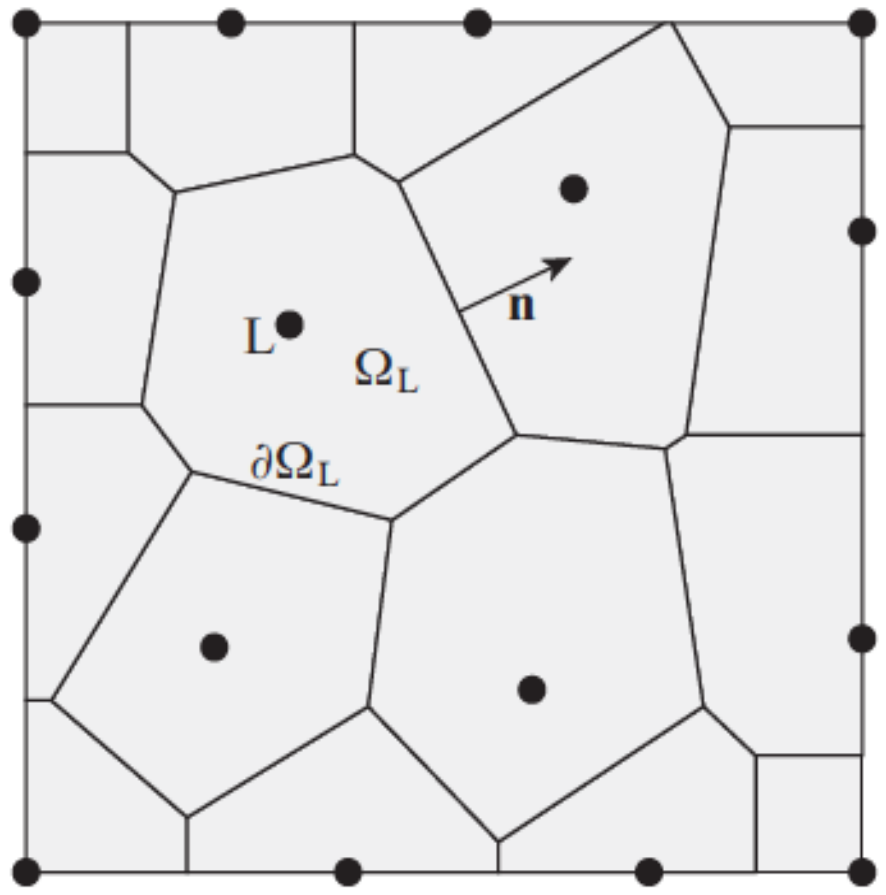

**Fig. 21.** Illustration of Voronoi diagram for SCNI.